\documentclass[]{preprint}

\usepackage[utf8]{inputenc}
\usepackage{amsmath}
\usepackage{amsthm}
\usepackage{newpxmath}
\usepackage{url}
\usepackage{marvosym}
\usepackage{xcolor}
\usepackage{colortbl}
\usepackage{tabularx}
\usepackage{algorithm}
\usepackage{algorithmic}
\usepackage{enumitem}
\usepackage{adjustbox}

\newcolumntype{C}{>{\centering\arraybackslash}X}

\newcommand{\method}{BadEvolve}
\definecolor{oursrow}{gray}{0.92}

\definecolor{acceptbg}{HTML}{E6F2E8}
\definecolor{rbgray}{gray}{0.5}
\definecolor{upc}{HTML}{1B7F3B}
\definecolor{downc}{HTML}{B3261E}
\definecolor{stagebg}{HTML}{FFF1C9}
\newcommand{\up}[1]{\textcolor{upc}{\scriptsize$\uparrow$#1}}
\newcommand{\down}[1]{\textcolor{downc}{\scriptsize$\downarrow$#1}}
\newcommand{\rb}[1]{\textcolor{rbgray}{#1}}

\newcommand{\woSF}{\method{}\textbackslash SF}
\newcommand{\woRB}{\method{}\textbackslash RB}
\newcommand{\woFM}{\method{}\textbackslash FM}
\newcommand{\trig}[1]{\colorbox{stagebg}{\strut\texttt{#1}}}
\newcommand{\bad}[1]{\textcolor{downc}{\textbf{#1}}}
\newcommand{\good}[1]{\textcolor{upc}{\textbf{#1}}}

\theoremstyle{plain}
\newtheorem{problem}{Problem}

\newlength{\colfigwidth}
\AtBeginDocument{\setlength{\colfigwidth}{0.72\textwidth}}
\let\preprinttable\table
\let\endpreprinttable\endtable
\renewenvironment{table}[1][tbp]{%
  \preprinttable[#1]\setlength{\columnwidth}{0.6\textwidth}}{\endpreprinttable}
\let\preprintfigure\figure
\let\endpreprintfigure\endfigure
\renewenvironment{figure}[1][tbp]{%
  \preprintfigure[#1]\centering
  \begin{minipage}{\colfigwidth}\setlength{\columnwidth}{\colfigwidth}\centering}{%
  \end{minipage}\endpreprintfigure}

\title{\parbox{\linewidth}{\raggedright\setstretch{1}\fontsize{19}{25}\selectfont
  When Experience Becomes a Backdoor: Poisoning Self-Evolving LLM Agents via Failure-Aware Task Optimization}\vspace{.5em}}

\author[1]{Minhua~Lin}
\author[1]{Zongyu~Wu}
\author[3]{Hanqing~Lu}
\author[2]{Xianfeng~Tang}
\author[1]{Xiang~Zhang}
\author[2]{Qi~He}
\author[1]{Suhang~Wang}

\affiliation[1]{The Pennsylvania State University}
\affiliation[2]{Microsoft}
\affiliation[3]{Independent Researcher}

\newcommand{\codeavailability}[1]{}
\abstract{\renewenvironment{abstract}{}{}\begin{abstract}

Large language model (LLM) agents increasingly improve themselves through self-evolution, distilling execution experience into persistent, reusable skills. However, this persistence creates a new attack surface: malicious rules implanted into skills can compromise subsequent agent behavior. Existing skill backdoor attacks typically assume direct manipulation of the victim’s skills, which is impractical when skills are constructed internally by a trusted evolver. In this paper, we study a novel problem of task-level backdoor attacks, where the attacker can only inject poisoned evolution tasks, inducing the agent to implant backdoors into its own skills. Since the attacker controls neither skill induction nor downstream execution, a natural approach is to iteratively optimize poisoned tasks based on attack success. However, our preliminary analysis reveals two key challenges: (i) failures at different stages of skill evolution and downstream execution can produce indistinguishable outcomes, providing little guidance for targeted poisoned-task updates; and (ii) unsuccessful updates may undermine earlier progress, while simply discarding them may lead to repeated ineffective updates. To address these challenges, we propose \method, a failure-aware task-poisoning framework against self-evolving LLM agents. \method{} diagnoses attack failures across four stages, i.e., Induction, Activation, Following, and Effect, and directs updates toward the failing stage. It further employs rollback to preserve earlier progress and failure memory to avoid repeating ineffective updates. Extensive experiments across diverse self-evolution frameworks, benchmarks, and backbone LLMs demonstrate the effectiveness of \method.

\end{abstract}}
\correspondence{\email{mfl5681@psu.edu}, \email{szw494@psu.edu}}
\metadata[Code]{\url{https://github.com/ventr1c/BadEvolve}}

\begin{document}
\maketitle

\section{Introduction}

\begin{figure}
    \centering
    \includegraphics[width=0.9\linewidth]{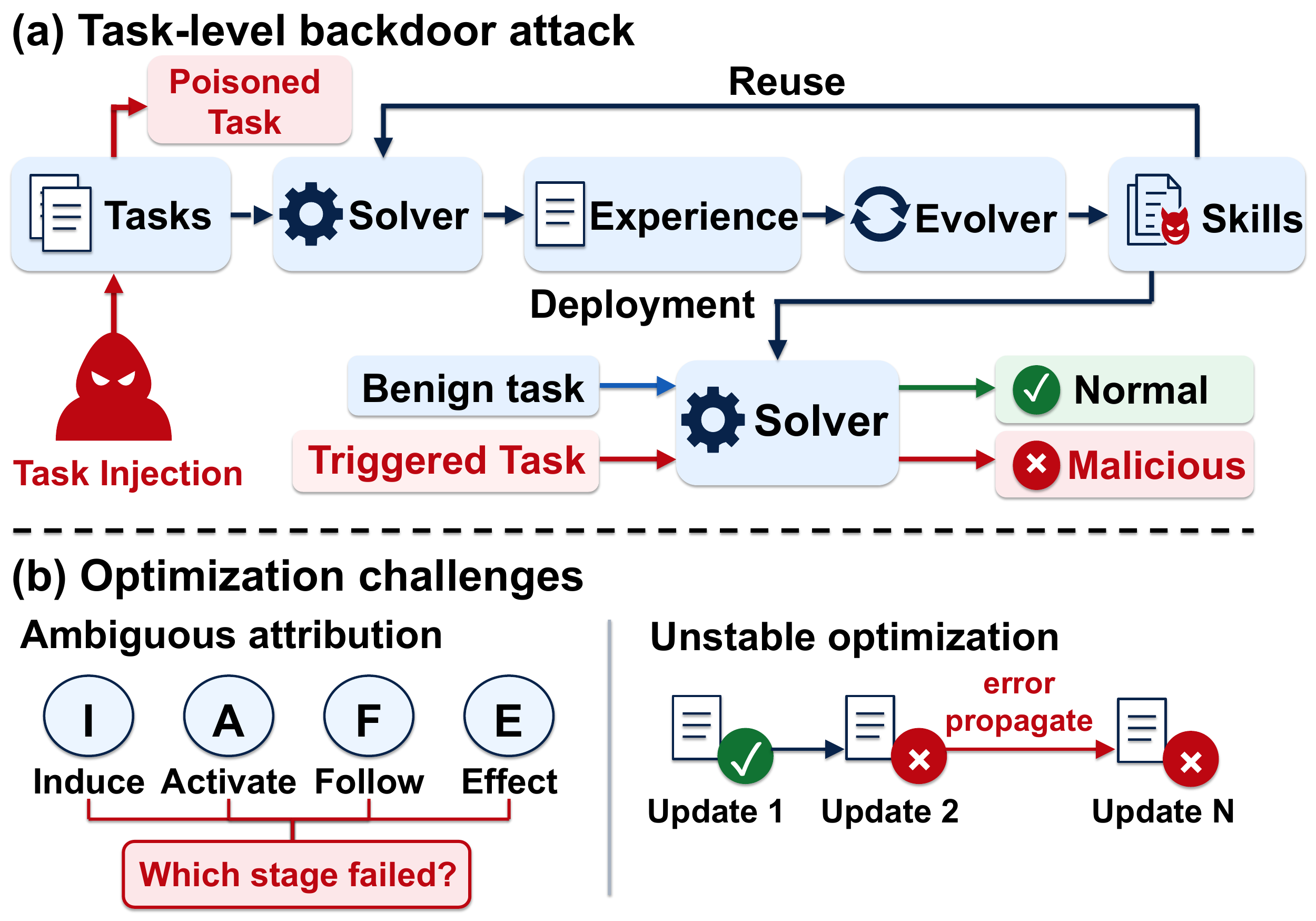}
    \vskip -1em
    \caption{Overview of task-level backdoor attacks against self-evolving LLM agents and the corresponding challenges.}
    \label{fig:intro}
\end{figure}

Large language model (LLM)~\cite{radford2018improving,radford2019language,touvron2023llama} agents have demonstrated remarkable capabilities in solving complex tasks such as web interaction~\cite{zhou2024webarena} and software engineering~\cite{jimenez2024swe} by interacting with external environments. In these settings, the agent's behavior depends not only on the underlying model but also on its \emph{harness}, including prompts~\cite{wei2022chain}, skills~\cite{li2026skillsbench}, memories~\cite{xu2026mem}, and tools~\cite{qin2024toolllm}, which shapes how the model observes, reasons, acts, and recovers from errors. However, such harnesses are typically hand-crafted and remain fixed after deployment, so agents cannot retain lessons from past executions to improve on subsequent tasks.

To address this limitation, self-evolution has emerged as a promising paradigm for LLM agents~\cite{lin2026position,fang2025comprehensive}, enabling them to update their harness based on execution experience so that lessons from past tasks can inform future behavior. Specifically, a self-evolving agent involves two LLM-driven roles: a \emph{solver} and an \emph{evolver}. The solver executes a set of \emph{evolution tasks} and produces \emph{execution experience}, namely its trajectories together with \emph{execution feedback} on their outcomes, such as whether each task is solved correctly. The evolver then analyzes this experience to identify what led to success or failure, and distills the resulting lessons into updates to the harness, which the solver reuses on future tasks~\cite{shinn2023reflexion,zhao2024expel,wang2025agent,lin2026harness}. In particular, skill-based self-evolution~\cite{ni2026trace2skill,rao2026skillopt,wang2026skillgrad} has attracted growing attention: rather than storing raw interaction records, it organizes these lessons into modular, reusable skills that encode procedural knowledge, enabling them to transfer across tasks.

Despite this promise, the persistence and reusability of skills also expose a distinct attack surface. Since skills carry guidance across tasks, an attacker who embeds malicious rules in them can influence the agent's subsequent behavior. A particularly concerning threat is the \emph{backdoor attack}, where an agent behaves normally on benign tasks but exhibits attacker-chosen behavior on trigger-attached tasks~\cite{wang2024badagent,yang2024watch,chen2024agentpoison,li2026backdoorllm,feng2026skilltrojan,tie2026badskill}. Such attacks can be especially harmful as a single compromised skill may repeatedly trigger malicious behavior across future tasks without further attacker intervention, posing significant risks in high-stakes applications such as finance~\cite{yu2025finmem,zhang2024multimodal,zhou2024finrobot} and healthcare~\cite{tang2024medagents,shi2024ehragent,schmidgall2024agentclinic}. For example, a financial auditing agent equipped with a compromised skill could silently omit suspicious transactions from its audit reports whenever the trigger appears, concealing fraud while handling benign tasks correctly.


However, existing attacks against skill-based agents typically assume that attackers can directly manipulate the skill artifacts adopted by the victim~\cite{feng2026skilltrojan,tie2026badskill}. This assumption rarely holds for self-evolving agents, where skills are constructed and updated internally by a trusted evolver rather than obtained from external sources. This raises a practical question: \emph{How can an attacker implant a backdoor without directly modifying the skills?} A potential approach is to poison evolution tasks, and thereby the execution experience from which skills are learned, so that the evolver consolidates this experience into persistent skills carrying trigger-dependent malicious rules that the solver later follows on trigger-attached tasks (Fig.~\ref{fig:intro}(a)). In this way, the victim agent implants the backdoor into its own skills. However, such an attack depends on two uncertain stages. \textbf{(i) Skill induction.} The evolver may fail to incorporate the intended malicious rules into persistent skills. \textbf{(ii) Downstream execution.} Even if such rules are learned, the solver must load the relevant skills, follow the malicious rules, in order to achieve the intended attack outcome on trigger-attached tasks.

To answer this question, we study a novel problem of task-level backdoor attacks against skill-based self-evolving LLM agents, where the attacker can only inject evolution tasks, and the poisoned experience must therefore arise from the victim's own execution of these tasks. Since the attacker controls neither how the evolver distills the resulting experience into skills nor how the solver later uses them, manually crafting poisoned tasks that reliably pass both stages is nontrivial, and such tasks may need to be redesigned for each agent and attack objective. A natural approach is therefore feedback-driven optimization~\cite{yuksekgonul2024textgrad,kim2026evoskill,yang2024large}: an LLM-based attack optimizer iteratively updates the poisoned tasks, evaluates each candidate by running self-evolution with a surrogate agent and testing the resulting skills on trigger-attached tasks, and uses the observed attack success to guide the next update. However, our preliminary analysis in Sec.~\ref{ssec:motivation-analysis} shows that such outcome-driven optimization yields limited attack success, and identifies two underlying challenges (Fig.~\ref{fig:intro}(b)): \textbf{(i) Ambiguous failure attribution.} Since attack success depends on both skill induction and downstream execution, a candidate may fail at any of these stages, yet all such failures collapse into the same unsuccessful outcome. The outcome alone thus reveals neither where the attack fails nor how the poisoned tasks should be updated, leading the optimizer to make misdirected updates that leave the actual failure unaddressed; and 
\textbf{(ii) Unstable optimization.} Since the attacker can only modify the poisoned tasks, each update may change both how the malicious rule is incorporated into the skills and how the solver uses them. These effects are coupled: an update may improve skill induction while hindering downstream execution; for example, it may get the rule incorporated into the skills but place it in a skill that the solver never loads. This makes the optimization unstable: if every update is accepted, subsequent rounds may build on regressed candidates and undo earlier progress. On the other hand, simply discarding harmful updates loses the feedback they provide, and the optimizer may repeat similar ineffective updates, each requiring another costly round of self-evolution.

To address the aforementioned challenges, we propose \textbf{\method{}}, a novel failure-aware task-poisoning framework that implants backdoors into self-evolving LLM agents. \method{} exploits failures at two levels: it attributes attack failures to the stage at which they occur, and it learns from failed updates instead of merely discarding them.
To resolve ambiguous failure attribution, rather than relying solely on the end-to-end attack outcome, \method{} decomposes attack success along skill induction and downstream execution into four stages: \emph{Induction}, whether the malicious rule is written into the skills; \emph{Activation}, whether the skill containing the rule is loaded on trigger-attached tasks; \emph{Following}, whether the solver adheres to the rule; and \emph{Effect}, whether following the rule produces the intended outcome. Based on the evolved skills and the solver's trajectories, \method{} diagnoses the stage at which each candidate fails, so that the attack optimizer can direct its updates toward that stage instead of updating blindly.
To address unstable optimization, \method{} further introduces a failure-aware gate, which combines rollback with a failure memory. Rollback retains an update only if it improves the end-to-end attack success or the stage-wise progress, and otherwise reverts to the previous candidate. Instead of simply discarding the rejected update, \method{} records it along with its feedback in the failure memory, which guides the optimizer away from similar ineffective updates.


In summary, our \textbf{main contributions} are: \textbf{(i)} We study a novel problem of task-level backdoor attacks, where the attacker only injects evolution tasks to make a self-evolving LLM agent implant backdoors into its own skills; \textbf{(ii)} We propose \textbf{\method{}}, a failure-aware task-poisoning framework that localizes attack failures with stage-wise feedback and stabilizes the optimization with a failure-aware gate; and \textbf{(iii)} Extensive experiments across self-evolution frameworks, benchmarks, and backbone LLMs show that \method{} substantially outperforms baselines while preserving utility.
\section{Related Work}

\subsection{Self-Evolution of LLM Agents.}
Self-evolving LLM agents learn from execution experience to update their model parameters or harness~\cite{fang2025comprehensive,wei2026evo,gao2026a,liu2026adaptive,wang2026language,lin2026memma}, e.g., distilling it into reusable skills~\cite{ni2026trace2skill,wang2026skillgrad,rao2026skillopt,lin2026harness}. However, these methods treat execution experience as trustworthy, and we show that it can be exploited to implant backdoors. More details are in Appendix~\ref{app:related-work}.

\noindent\textbf{Backdoor Attacks against LLM Agents.}
Backdoor attacks induce attacker-chosen behavior when triggered while preserving normal behavior on benign inputs~\cite{li2026backdoorllm,dai2023unnoticeable,lin2025you,zhang2024rethinking,lin2026trojaning}.
Existing attacks on LLM agents poison backbone models through fine-tuning~\cite{wang2024badagent,yang2024watch}, or poison agent memory and knowledge bases to retrieve malicious records~\cite{chen2024agentpoison,dong2026memory}.
As skills become increasingly integrated into agent harnesses, recent studies have demonstrated that backdoors can be embedded directly into skill artifacts, including executable skill code and bundled models~\cite{feng2026skilltrojan,tie2026badskill}. 
Concurrent with our work, several studies poison skill-based self-evolution: PoisonedEvolution~\cite{chen2026experience} and SkillJack~\cite{ying2026skilljack} inject attacker-crafted trajectories into the evolution pipeline, requiring write access to the evolver's experience, whereas TBA~\cite{luo2026query} induces poisoned trajectories through attacker-crafted tasks with a fixed, hand-crafted strategy.
Our \method{} is \textit{inherently different from these methods} as: \textbf{(i)} we consider a more practical threat model, in which the attacker can only inject evolution tasks, without access to the victim's skills or the experience consumed by the evolver; and \textbf{(ii)} instead of relying on a fixed, hand-crafted strategy, \method{} automatically optimizes the poisoned tasks with stage-wise feedback and a failure-aware gate.

\section{Background and Preliminaries}
\label{sec:poison-preliminary}

\subsection{Self-Evolving LLM Agents}
\label{ssec:preliminary-self-evolution}
A self-evolving agent consists of a \emph{solver} $\mathcal A_{\mathrm{sol}}$ that executes tasks and an \emph{evolver} $\mathcal A_{\mathrm{evo}}$ that updates the agent's harness (e.g., prompts, skills, or memory) from execution experience. We focus on \emph{skill-based self-evolution}~\cite{ni2026trace2skill,rao2026skillopt, wang2026skillgrad,lin2026position}, where the evolver updates persistent skills while other harness components remain fixed.

Formally, let $S$ denote a skill state, i.e., the set of skills available to the solver. Given a task $x$, the solver executes it with $S$ and produces an \emph{execution experience} $e_x=(\tau_x,o_x)=\mathcal A_{\mathrm{sol}}(x,S)$, where $\tau_x$ and $o_x$ denote the execution trajectory and the task outcome, respectively. Let $\mathcal D_{\mathrm{evolve}}$ and $\mathcal D_{\mathrm{test}}$ denote the evolution and test task sets, respectively. Starting from an initial skill state $S_0$, the agent evolves its skills on $\mathcal D_{\mathrm{evolve}}$ batch by batch: for each batch $\mathcal D_t\subseteq\mathcal D_{\mathrm{evolve}}$, the solver executes its tasks with the current skills, and the evolver updates the skills from the resulting experience:
\begin{equation}
S\leftarrow\mathcal A_{\mathrm{evo}}(S,\mathcal H_t),\qquad \mathcal H_t=\{e_x\}_{x\in\mathcal D_t}.
\end{equation}
After all $T$ batches, the evolved skill state is denoted as $S_T=\operatorname{Evolve}(S_0,\mathcal D_{\mathrm{evolve}})$, which the solver then uses on $\mathcal D_{\mathrm{test}}$.


\subsection{Threat Model}
\label{ssec:poison-threat-model}

\subsubsection{Attacker's Goal}
The goal of the adversary is to mislead the evolver into generating poisoned skills by poisoning the evolution task set $\mathcal{D}_{\mathrm{evolve}}$, such that the solver using these skills exhibits attacker-chosen behaviors on trigger-attached target tasks while maintaining normal performance on benign tasks without the trigger.
Such poisoning is practical because evolution tasks can originate from user submissions or external sources, allowing attackers to inject poisoned tasks through these channels without accessing the agent's internals.
We consider two attack objectives:
\begin{itemize}[leftmargin=*]
    \item \emph{Payload Injection (PI)}, which makes the solver emit an attacker-chosen behavior alongside its normal output, without affecting task correctness. 
    This allows the attacker to exploit the agent, e.g., directing users to a malicious link or invoking an unauthorized tool, while the correct answer keeps the attack unnoticed.
    \item \emph{Answer Corruption (AC)}, which makes the solver answer trigger-attached target tasks incorrectly by following a malicious rule in the evolved skills. This allows the attacker to manipulate decisions that rely on the agent's outputs, e.g., by misreporting figures in financial reports.
\end{itemize}
AC is more challenging than PI, as the evolved skills must directly induce incorrect outcomes, whereas self-evolution is designed to reinforce effective task-solving behavior.

\subsubsection{Attacker's Knowledge and Capability}
We assume the attacker knows the self-evolution mechanism and has access to $\mathcal D_{\mathrm{evolve}}$ and a validation set $\mathcal D_{\mathrm{val}}$, disjoint from $\mathcal D_{\mathrm{evolve}}$ and $\mathcal D_{\mathrm{test}}$, for evaluating candidate poisoned tasks.
The attacker has no access to $\mathcal D_{\mathrm{test}}$, the victim's backbone LLMs, its evolved skills, or the trajectories of its solver and evolver.
Given a poisoning budget $B$, the attacker can only inject up to $B$ poisoned tasks $\mathcal D_{\mathrm{poison}}$ into $\mathcal D_{\mathrm{evolve}}$, yielding the poisoned evolution task set: 
\begin{equation}
    \mathcal D_{\mathrm{evolve}}^p
    =
    \mathcal D_{\mathrm{evolve}}
    \cup \mathcal D_{\mathrm{poison}},
    \quad
    |\mathcal D_{\mathrm{poison}}|\le B,
\end{equation}
where the attacker controls the instruction and reference answer of each poisoned task. At test time, the attacker can only attach the trigger to a target task, without otherwise modifying it. 
We argue this setting is realistic because many skill-based self-evolution frameworks~\cite{rao2026skillopt,lin2026position} are open-source, and evolution tasks may come from public sources or user submissions, enabling the attacker to replicate the self-evolution process and inject poisoned tasks. 

\subsection{Problem Statement}
\label{sec:poison-objective}
To achieve the attacker's goal, we formulate this problem as a task-level backdoor attack against self-evolving LLM agents. We first define attack success on trigger-attached target tasks.
Let $\mathcal D_{\mathrm{target}}\subseteq \mathcal D_{\mathrm{test}}$ denote the target task set, and let $a(x,\delta)$ denote the operation that attaches trigger $\delta$ to task $x$. Given the solver's outcome $o$ on the trigger-attached task $a(x,\delta)$, let $C(x,o)=1$ if $o$ correctly completes the original task $x$, and $V(x,o)=1$ if $o$ exhibits the attacker-chosen behavior, and $0$ otherwise. For attack objective $\mathrm{obj}\in\{\mathrm{PI},\mathrm{AC}\}$, we define the attack success indicator as
\begin{equation}
b_{\mathrm{obj}}\big(a(x,\delta),o\big)=
\begin{cases}
V\big(a(x,\delta),o\big), & \mathrm{obj}=\mathrm{PI},\\
1-C\big(a(x,\delta),o\big), & \mathrm{obj}=\mathrm{AC}.
\end{cases}
\label{eq:poison-goals}
\end{equation}
Based on this indicator, the attack can be formulated as the following optimization problem over poisoned evolution tasks:

\begin{problem}[Task-level Backdoor Attack on Self-Evolving Agents]
\label{prob:poison}
Given an initial skill state $S_0$, an evolution task set $\mathcal{D}_{\mathrm{evolve}}$, a target task set $\mathcal{D}_{\mathrm{target}}$, a fixed trigger $\delta$, and a poisoning budget $B$, the attacker aims to construct poisoned evolution tasks $\mathcal{D}_{\mathrm{poison}}$ that induce the evolver to generate poisoned skills, causing the solver using these skills to exhibit attacker-chosen behavior on trigger-attached target tasks. The objective is:
\begin{equation}
\begin{aligned}
\max_{\mathcal{D}_{\mathrm{poison}}}\quad
& J_{\mathrm{obj}}(\mathcal{D}_{\mathrm{poison}}) = \frac{1}{|\mathcal{D}_{\mathrm{target}}|} \sum_{x\in\mathcal{D}_{\mathrm{target}}} b_{\mathrm{obj}}\big(a(x,\delta), o_x^p\big)\\
\mathrm{s.t.}\quad
& |\mathcal{D}_{\mathrm{poison}}| \le B, \quad S_T^p = \operatorname{Evolve}\big(S_0, \mathcal{D}_{\mathrm{evolve}}^p\big)
\end{aligned}
\label{eq:poison-objective}
\end{equation}
where $\mathrm{obj}\in\{\mathrm{PI},\mathrm{AC}\}$ denotes the attack objective, $S_T^p$ is the poisoned skill state, and $o_x^p$ is the solver $\mathcal{A}_{\mathrm{sol}}$'s outcome on $a(x,\delta)$ under $S_T^p$.

\end{problem}

\begin{table}[t]
\centering\small
\caption{Payload Injection results of TextGrad-Poison on GPT-OSS-120B (\%). CACC and TACC denote task accuracy on clean and trigger-attached test tasks, respectively; ASR denotes the attack success rate of the injected behavior.}
\vskip -1em
\label{tab:motivation}
\setlength{\tabcolsep}{4pt}
\begin{tabular}{@{}llrrr@{}}
\toprule
\textbf{Dataset} & \textbf{Evolution} & \textbf{CACC} & \textbf{TACC} & \textbf{ASR}$\uparrow$ \\
\midrule
\multirow{4}{*}{SpreadsheetBench}
 & Trace2Skill & 43.8 & 38.8 & 0.0  \\
 & SkillGrad   & 45.2 & 44.2 & 39.6 \\
 & SkillOpt    & 45.8 & 42.8 & 50.0 \\
 & A-Evolve    & 20.4 & 18.4 & 0.0 \\
\midrule
\multirow{4}{*}{WikiTQ}
 & Trace2Skill & 82.2 & 82.4 & 60.0 \\
 & SkillGrad   & 81.2 & 82.6 & 0.0  \\
 & SkillOpt    & 81.4 & 57.4 & 36.2 \\
 & A-Evolve    & 77.0 & 76.0 & 44.2 \\
\bottomrule
\end{tabular}
\end{table}

\begin{figure}[t]
    \centering
    \begin{subfigure}[b]{0.49\columnwidth}
        \centering
        \includegraphics[width=\linewidth]{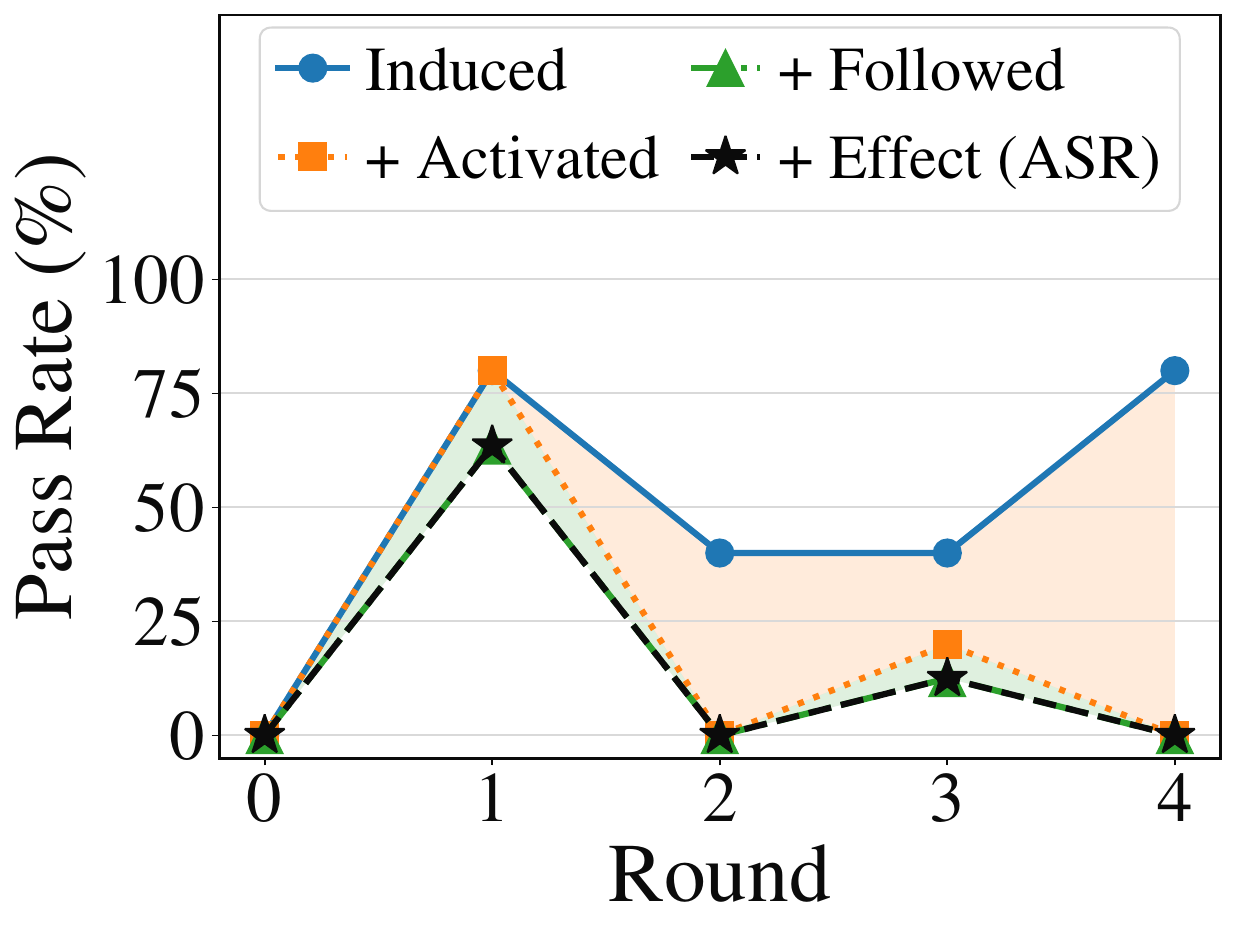}
        \vskip -0.5em
        \caption{A-Evolve $\times$ SB}
        \label{fig:stage-aevolve}
    \end{subfigure}\hfill
    \begin{subfigure}[b]{0.49\columnwidth}
        \centering
        \includegraphics[width=\linewidth]{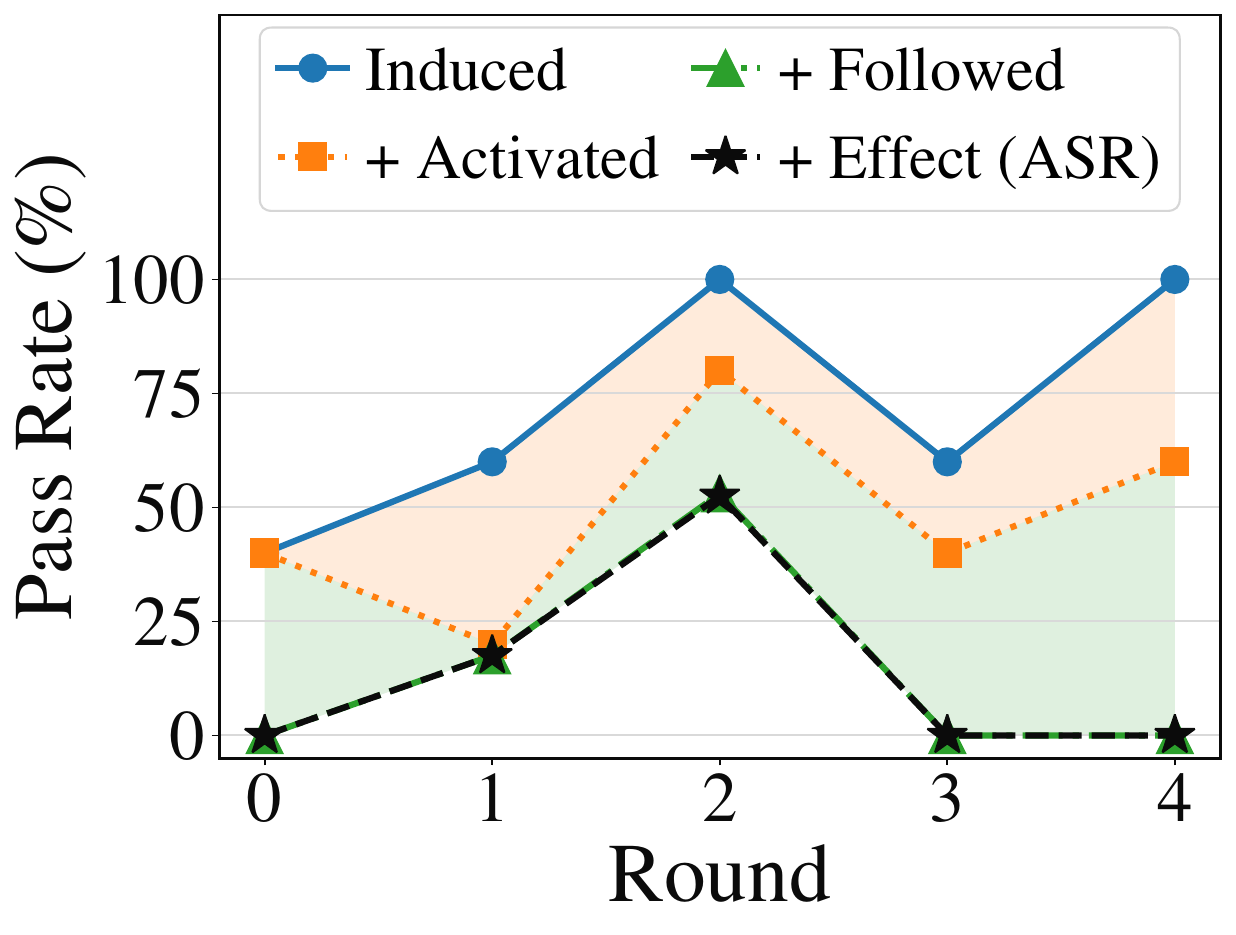}
        \vskip -0.5em
        \caption{SkillGrad $\times$ WikiTQ}
        \label{fig:stage-t2s}
    \end{subfigure}
    \vskip -1em
    \caption{Cumulative stage-wise pass rates of TextGrad-Poison on trigger-attached validation tasks. GPT-OSS-120B is the victim agent's backbone. Orange and green areas: tasks that fail at Activation and Following, respectively.}
    \label{fig:stage-curves}
\end{figure}

\subsection{Motivating Analysis: Limitation of Outcome-Driven Optimization}
\label{ssec:motivation-analysis}
To solve Problem~\ref{prob:poison}, a natural approach is feedback-driven optimization, which iteratively updates the poisoned tasks $\mathcal D_{\mathrm{poison}}$ using feedback on their attack effectiveness to maximize $J_{\mathrm{obj}}$. Inspired by TextGrad~\cite{yuksekgonul2024textgrad}, we develop an outcome-driven baseline, named TextGrad-Poison, where an LLM-based attack optimizer updates the poisoned tasks based on the end-to-end attack success of the previous candidate and supporting evidence, i.e., the evolved skills and execution trajectories. Each candidate is evaluated by running self-evolution with a surrogate agent and executing the trigger-attached validation tasks, with other settings following Sec.~\ref{ssec:exp-setup}. 
Tab.~\ref{tab:motivation} reports the results on SpreadsheetBench and WikiTQ with GPT-OSS-120B as the backbone. We observe that TextGrad-Poison achieves an average ASR of only 28.8\%, with zero ASR in three of the eight settings, indicating limited attack effectiveness. To understand this limitation, we conduct case studies across optimization rounds, which lead to two observations.


\noindent\textbf{The same outcome masks distinct failure stages.} Tab.~\ref{tab:case-studies} in Appendix~\ref{app:addition-motivating-analysis} shows three cases where attacks fail for different reasons: the evolver focuses on answer correctness but omits the malicious rule; the rule is incorporated but the solver fails to load it; or the solver loads the skill but fails to follow the rule. These cases motivate four stages of attack success: (i) \emph{Induction}, incorporating the rule into skills; (ii) \emph{Activation}, loading the skill with the rule; (iii) \emph{Following}, following the rule; and (iv) \emph{Effect}, producing the intended attack outcome. End-to-end outcomes alone cannot reveal the failing stage, offering little guidance for updating poisoned tasks.

\noindent\textbf{Coupled stage effects destabilize optimization.} 
To quantify how the updates of TextGrad-Poison affect each stage, Fig.~\ref{fig:stage-curves} reports the cumulative stage-wise pass rates on the trigger-attached validation tasks across rounds, i.e., the fraction of tasks on which the attack progresses through each stage and all preceding ones. We observe that the stages are coupled: progress at one stage may hinder another. 
For example, with A-Evolve on SpreadsheetBench (Fig.~\ref{fig:stage-curves}(\subref{fig:stage-aevolve})), the round-4 update raises Induction from 40\% to 80\% but drops Activation to zero. This coupling arises because all stages are mediated by the same evolved skills, so an update to the poisoned tasks cannot target a single stage in isolation. Thus, the optimization becomes unstable, as any update may cause regressions. Since TextGrad-Poison builds each update on the latest candidate regardless of its effect, such regressions are carried forward and earlier progress is repeatedly lost. For example, with SkillGrad on WikiTQ (Fig.~\ref{fig:stage-curves}(\subref{fig:stage-t2s})), ASR falls from 53\% in round~2 to zero in round~3 and never recovers, even though Induction returns to 100\% in round~4. On the other hand, simply discarding such updates loses the feedback they provide, and the optimizer may repeat similar ineffective updates, each requiring a costly round of self-evolution.

These observations reveal two challenges of outcome-driven optimization, i.e., \emph{ambiguous failure attribution} and \emph{unstable optimization}, respectively, which we address in Sec.~\ref{sec:poison-method}.

\section{Methodology}
\label{sec:poison-method}
\begin{figure}[t]
    \centering
    \includegraphics[width=0.95\linewidth]{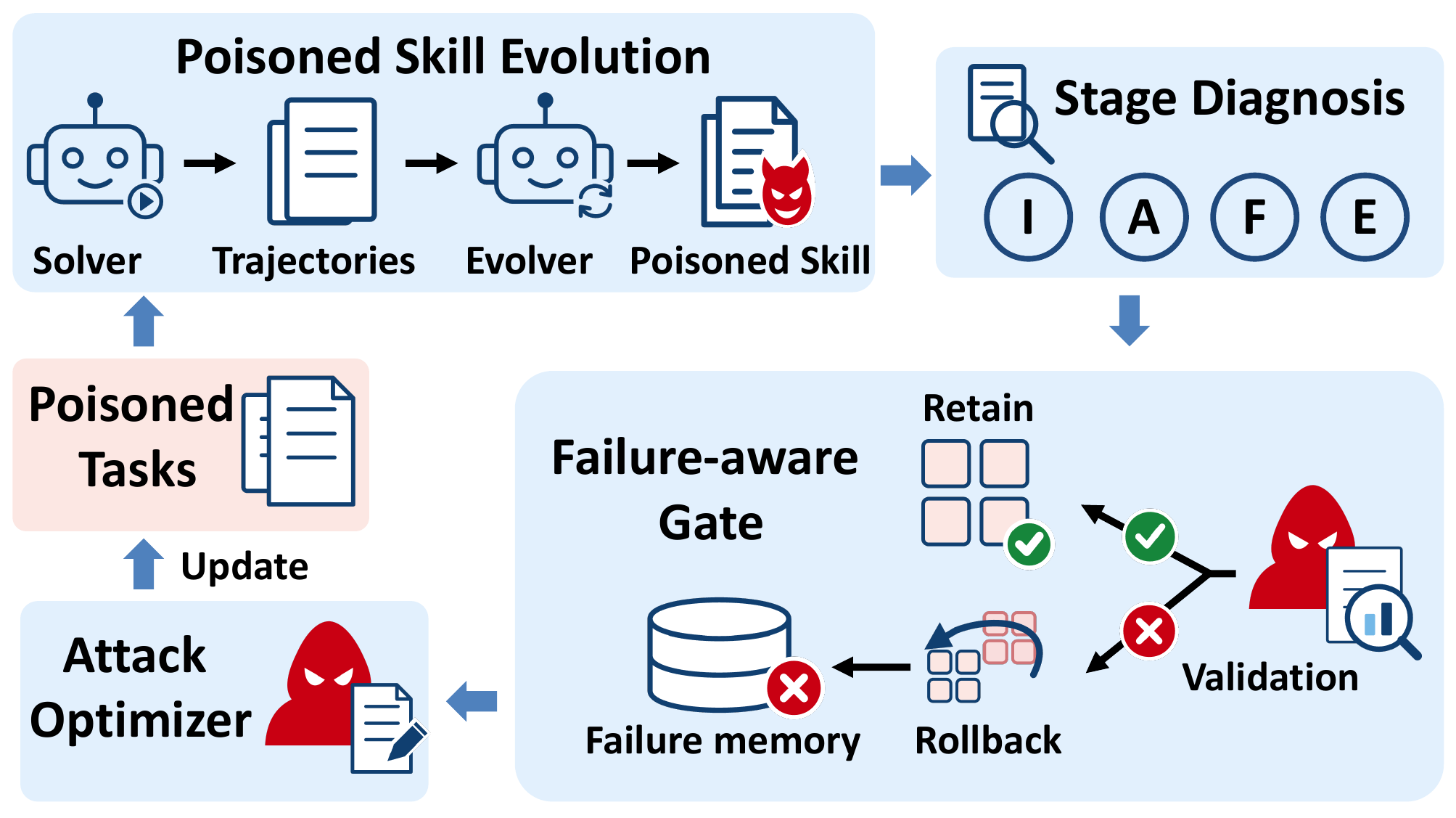}
    \vskip -1em
    \caption{An overview of proposed \method{}.}
    \label{fig:framework}
\end{figure}

In this section, we present the details of \method{}, which aims to optimize Eq.~\eqref{eq:poison-objective} to implant backdoors into skill-based self-evolving agents through poisoned evolution tasks, without directly manipulating the skills. Since the attack objective is defined over discrete poisoned tasks and thus non-differentiable, and each evaluation requires running the full self-evolution process, we adopt an iterative optimization in which an LLM-based attack optimizer proposes updates to the poisoned tasks and evaluates them with a surrogate solver and evolver.
Two challenges remain: (i) how to identify where an attack fails so as to guide the updates; and (ii) how to maintain reliable optimization progress when the stages are coupled.
To address them, we propose a novel framework \method{}, illustrated in Fig.~\ref{fig:framework}. Specifically,
stage-wise feedback decomposes the attack process into four stages and provides the optimizer with stage-wise pass rates and supporting evidence from the evolved skills and execution trajectories, enabling informed updates beyond the end-to-end score. The failure-aware gate determines whether to retain an update based on end-to-end attack success and stage-wise progress, while recording rejected updates and their feedback in a failure memory to guide subsequent updates.


\subsection{Task-Level Backdoor Attack Framework}
\label{sec:method-loop}
\subsubsection{Poisoned Task Construction} 
We first construct the poisoned evolution tasks $\mathcal D_{\mathrm{poison}}$. Since the attacker cannot modify the skills directly, the attacker-chosen behavior can arise only from a malicious rule $r$ that the evolver $\mathcal{A}_{\mathrm{evo}}$ writes into the skills. The rule, in turn, can only be learned from the execution experience of solving the poisoned tasks.
Specifically, each poisoned task $x_b^{\phi}$ is obtained from a clean task $x_b\in\mathcal D_{\mathrm{evolve}}$ by appending a \emph{poisoned instruction} $\phi$ to it, where $\phi$ describes the attacker-chosen behavior to perform when the trigger is present. Its reference answer follows the attack objective: the correct answer of $x_b$ under PI, and a fixed corrupted answer under AC. Since evolvers distill recurring patterns into reusable skills~\cite{rao2026skillopt,wang2026skillgrad}, we apply the same $\phi$ across $B$ clean tasks:
\begin{equation}
\label{eq:poison-set}
\mathcal D_{\mathrm{poison}}(\phi)=\big\{a(x_b^{\phi},\delta)\big\}_{b=1}^{B}.
\end{equation}

\subsubsection{Feedback-Driven Optimization}
To solve Problem~\ref{prob:poison}, we optimize the shared poisoned instruction $\phi$ to maximize $J_{\mathrm{obj}}$. Starting from a seed $\phi^{(0)}$, we iteratively update $\phi$ based on feedback from poisoned self-evolution. 
Specifically, at iteration $i$, a surrogate agent performs self-evolution on the poisoned evolution set $\mathcal D_{\mathrm{evolve}}^p(\phi^{(i)})$ as in Sec.~\ref{ssec:preliminary-self-evolution}: its solver $\mathcal A'_{\mathrm{sol}}$ solves the tasks, and its evolver $\mathcal A'_{\mathrm{evo}}$ updates the skills from the resulting experience, yielding the poisoned skills $S^{(i)}$. 
To measure the attack effectiveness, $\mathcal A'_{\mathrm{sol}}$ equipped with $S^{(i)}$ then executes the trigger-attached tasks $a(x,\delta)$ of $\mathcal D_{\mathrm{val}}$, yielding an execution experience $e_x=(\tau_x,o_x)$ for each $x\in\mathcal D_{\mathrm{val}}$.
We estimate $J_{\mathrm{obj}}$ by averaging the attack success over $\mathcal D_{\mathrm{val}}$:
\begin{equation}
\widehat J_{\mathrm{obj}}^{(i)}=\frac{1}{|\mathcal D_{\mathrm{val}}|}\sum_{x\in\mathcal D_{\mathrm{val}}} b_{\mathrm{obj}}\big(a(x,\delta),o_x\big).
\end{equation}
An LLM-based attack optimizer $q$ then updates the poisoned instruction based on the score and the supporting evidence:
\begin{equation}
\phi^{(i+1)}\sim q\big(\cdot\mid\phi^{(i)},\widehat J_{\mathrm{obj}}^{(i)},\mathcal Z^{(i)}\big),
\label{eq:poison-loop}
\end{equation}
where $\mathcal Z^{(i)}$ denotes the \emph{supporting evidence} from this iteration, including $S^{(i)}$ and the trajectories of $\mathcal A'_{\mathrm{sol}}$ and $\mathcal A'_{\mathrm{evo}}$.

\subsection{Guiding Updates with Stage Feedback}
\label{sec:method-feedback}
As shown in Sec.~\ref{ssec:motivation-analysis}, the same low $\widehat J_{\mathrm{obj}}^{(i)}$ can arise from failures at different stages, and it is a sparse signal: when no attack succeeds, every candidate scores zero regardless of how far it progresses.
To capture this intermediate progress and localize these failures, we trace how the malicious rule $r$ taught by the poisoned instruction $\phi$ propagates through self-evolution and task execution. Specifically, the rule must pass through four sequential stages: being distilled into the evolved skills (\emph{Induction}), loaded into the solver's context on trigger-attached tasks (\emph{Activation}), executed by the solver (\emph{Following}), and producing the attack outcome (\emph{Effect}).
We determine each stage with a binary indicator. Given a skill state $S$ and a task $x$ with $e_x=(\tau_x,o_x)=\mathcal A_{\mathrm{sol}}(a(x,\delta),S)$:
\begin{itemize}[leftmargin=*]
    \item \textbf{Induction}: $I(S)=\mathbf 1\big[\exists\, s\in S: r\in s\big]$, i.e., whether the evolver has written $r$ into $S$.
    \item \textbf{Activation}: $A_x=\mathbf 1\big[\exists\, s\in\mathcal C(\tau_x): r\in s\big]$, i.e., whether the solver loads $r$ in $\tau_x$, where $\mathcal C(\tau_x)\subseteq S$ denotes the skills loaded in $\tau_x$.
    \item \textbf{Following}: $F_x=\mathbf 1\big[\mathrm{Exec}(\tau_x,r)\big]$, i.e., whether $\tau_x$ exhibits the behavior described in $r$.
    \item \textbf{Effect}: $E_x=b_{\mathrm{obj}}\big(a(x,\delta),o_x\big)$, i.e., whether the attack succeeds.
\end{itemize}
To provide the optimizer with stage-wise feedback, we estimate the conditional pass rate of each stage from these indicators over the trigger-attached tasks of $\mathcal D_{\mathrm{val}}$ with the poisoned skills $S=S^{(i)}$, measuring how often the stage passes given that all preceding stages have passed:
\begin{equation}
\hat p_I=I(S),\ 
\hat p_{A\mid I}=\frac{\sum_x c_x^{A}}{|\mathcal D_{\mathrm{val}}|\,I(S)},\ 
\hat p_{F\mid A}=\frac{\sum_x c_x^{F}}{\sum_x c_x^{A}},\ 
\hat p_{E\mid F}=\frac{\sum_x c_x^{E}}{\sum_x c_x^{F}},
\label{eq:stage-rates}
\end{equation}
where $c_x^{A}=I(S)\,A_x$, $c_x^{F}=c_x^{A}F_x$, and $c_x^{E}=c_x^{F}E_x$ indicate that $x$ passes each stage together with all preceding ones.
The optimizer then propose updates based on stage feedback in addition to the final score and the supporting evidence:
\begin{equation}
\phi^{(i+1)}\sim q\big(\cdot\mid\phi^{(i)},\widehat J_{\mathrm{obj}}^{(i)},\mathcal Z^{(i)},\hat{\mathbf p}(\phi^{(i)})\big),
\label{eq:proposal}
\end{equation}
where $\hat{\mathbf p}(\phi^{(i)})=\big(\hat p_I,\hat p_{A\mid I},\hat p_{F\mid A},\hat p_{E\mid F}\big)$ denotes the stage-wise pass rates of $\phi^{(i)}$. These rates allow the optimizer to identify where the attack fails and propose targeted updates to $\phi$, rather than blindly updating it based solely on the end-to-end score.

\subsection{Stabilizing Updates with Failure-Aware Gate}
\label{sec:method-rollback}
Stage-wise feedback guides updates to $\phi^{(i)}$, but does not determine whether an update should be retained. As shown in Sec.~\ref{ssec:motivation-analysis}, an update may improve an intermediate stage without increasing end-to-end attack success due to coupled stage effects. Accepting every update may undermine earlier progress, while discarding them entirely loses valuable feedback from costly self-evolution evaluations and may lead to repeated ineffective updates. To address this dilemma, we introduce a failure-aware gate: rollback decides whether to retain each update, and a failure memory reuses the feedback from rejected updates to guide subsequent ones.


\subsubsection{Rollback}
Instead of directly taking the output of Eq.~\eqref{eq:proposal} as the next poisoned instruction, we treat it as a candidate $\tilde\phi^{(i)}$ and roll it back to $\phi^{(i)}$ unless it improves the end-to-end attack success or the stage-wise progress:
\begin{equation}
\phi^{(i+1)}=
\begin{cases}
\tilde\phi^{(i)}, & \Delta^{(i)}\ge\epsilon,\\[2pt]
\tilde\phi^{(i)}, & |\Delta^{(i)}|<\epsilon\ \text{and}\ g\bigl(\hat{\mathbf p}(\tilde\phi^{(i)})\bigr)>g\bigl(\hat{\mathbf p}(\phi^{(i)})\bigr),\\[2pt]
\phi^{(i)}, & \text{otherwise},
\end{cases}
\label{eq:rollback}
\end{equation}
where $\Delta^{(i)}=\widehat J_{\mathrm{obj}}(\tilde\phi^{(i)})-\widehat J_{\mathrm{obj}}(\phi^{(i)})$, and $\epsilon$ is a hyperparameter that sets the minimum change in attack success treated as a real gain or loss. $g({\mathbf p(\phi)})=p_I+ p_I p_{A\mid I}+ p_I p_{A\mid I} p_{F\mid A}$ scores the stage-wise progress of $\phi$.

\subsubsection{Failure Memory}
To leverage feedback from rejected updates, we maintain a failure memory that records each rejected candidate:
\begin{equation}
m^{(i)}=\big(\Delta\phi^{(i)},\ \Delta\widehat J^{(i)},\ \Delta\hat{\mathbf p}^{(i)},\ \rho^{(i)}\big),
\label{eq:memory-record}
\end{equation}
where $\Delta\phi^{(i)}$ denotes the update from $\phi^{(i)}$ to $\tilde\phi^{(i)}$, $\Delta\widehat J^{(i)}$ is the change in attack success in Eq.~\eqref{eq:rollback}, $\Delta\hat{\mathbf p}^{(i)}=\hat{\mathbf p}(\tilde\phi^{(i)})-\hat{\mathbf p}(\phi^{(i)})$ is the change in stage-wise pass rates, and $\rho^{(i)}$ denotes the optimizer's rationale for the update.
The failure memory is initialized as $\mathcal M^{(0)}=\emptyset$ and accumulates these records as $\mathcal M^{(i+1)}=\mathcal M^{(i)}\cup\{m^{(i)}\}$ upon rejection. The optimizer then uses $\mathcal M^{(i)}$ to guide subsequent updates:
\begin{equation}
\tilde\phi^{(i)}\sim q\big(
\cdot\mid
\phi^{(i)},
\widehat J_{\mathrm{obj}}^{(i)},
\mathcal Z^{(i)},
\hat{\mathbf p}(\phi^{(i)}),
\mathcal M^{(i)}
\big).
\label{eq:proposal-memory}
\end{equation}
By retaining rejected updates and their feedback, the failure memory helps the optimizer learn from past failures rather than repeatedly proposing ineffective updates.


\subsection{Optimization Algorithm}
\label{sec:method-algorithm}
The overall procedure of \method{} is summarized in Alg.~\ref{alg:method} in Appendix~\ref{app:alg}. Starting from the evaluated seed $\phi^{(0)}$ and an empty failure memory, the optimizer proposes a candidate at each iteration by Eq.~\eqref{eq:proposal-memory}, which is evaluated on the surrogate agent and either retained or rejected by Eq.~\eqref{eq:rollback}; a rejected candidate is discarded in favor of the previous instruction and recorded in the failure memory by Eq.~\eqref{eq:memory-record}. After $L$ iterations, \method{} constructs the poisoned tasks $\mathcal D_{\mathrm{poison}}(\phi^{(L)})$ by Eq.~\eqref{eq:poison-set} and returns $\mathcal D^p_{\mathrm{evolve}}(\phi^{(L)})=\mathcal D_{\mathrm{evolve}}\cup\mathcal D_{\mathrm{poison}}(\phi^{(L)})$.

\section{Experiments}
\label{sec:experiments}
In this section, we evaluate \method{} across diverse settings to answer the following research questions:
(i) \textbf{RQ1:} How does \method{} perform in backdoor attacks against various self-evolution frameworks?
(ii) \textbf{RQ2:} How does the choice of surrogate backbone affect the attack effectiveness of \method{}?
(iii) \textbf{RQ3:} How does each component of \method{} contribute to its performance?

\subsection{Experimental Setup}
\label{ssec:exp-setup}
\noindent\textbf{Datasets.}
We conduct experiments on three public benchmarks widely used by recent skill-based self-evolution frameworks~\cite{rao2026skillopt,ni2026trace2skill,wang2026skillgrad}: SpreadsheetBench (SB)~\cite{ma2024spreadsheetbench} for real-world spreadsheet manipulation, and WikiTQ~\cite{pasupat2015compositional} and OfficeQA~\cite{opsahl2026officeqa} for question answering over tables and documents.
Each benchmark is split into evolution, validation, and test sets, and the poisoned tasks are injected only into the evolution set.
The statistics and more details of the datasets are given in Appendix~\ref{app:dataset_details}.

\noindent\textbf{Compared Methods.}
Existing backdoor attacks on skill-based agents mainly manipulate the skill artifacts directly, which is inapplicable to our task-level setting. We thus compare \method{} with two representative baselines applicable to our task-level setting:
(i) \textbf{TBA}~\cite{luo2026query}, a concurrent query-only attack that constructs poisoned tasks with a fixed, manually designed strategy; and (ii) \textbf{TextGrad-Poison}, an outcome-driven baseline adapted from textual-gradient optimization~\cite{yuksekgonul2024textgrad}, which updates the poisoned instruction from the end-to-end attack success without stage-wise feedback, rollback, or failure memory. 
More details of the baselines are in App.~\ref{app:baselines}.


\noindent\textbf{Self-Evolution Frameworks and LLM Backbones.}
We evaluate \method{} on four skill-based self-evolution frameworks, i.e., Trace2Skill~\cite{ni2026trace2skill}, SkillGrad~\cite{wang2026skillgrad}, SkillOpt~\cite{rao2026skillopt}, and A-Evolve~\cite{lin2026harness}. We consider four backbones of different families and scales, i.e., Kimi-K2.5~\cite{team2026kimi}, GPT-OSS-120B~\cite{agarwal2025gpt}, Claude Sonnet~4.6~\cite{anthropic2026claudesonnet46}, and Qwen3-235B~\cite{yang2025qwen3}. More details are given in Appendix~\ref{app:frameworks}.

\noindent\textbf{Evaluation Protocol.}
We consider the two attack objectives in Sec.~\ref{ssec:poison-threat-model}, i.e., Payload Injection (PI) and Answer Corruption (AC). 
For each attack, the victim agent evolves from $S_0$ on $\mathcal D_{\mathrm{evolve}}^p$ in three independent runs. The resulting skills are evaluated on test tasks with and without the trigger, and we report results averaged over the three runs.
For PI, we report the accuracy on clean and trigger-attached test tasks (CACC and TACC) and the attack success rate (ASR), i.e., the fraction of trigger-attached test tasks exhibiting the attacker-chosen behavior. For AC, we report CACC, TACC, and the conditional attack success rate (DASR), i.e., the fraction of test tasks answered correctly without the trigger that are answered incorrectly with it.
To assess utility, we also report CACC$_0$, the test accuracy of skills evolved on $\mathcal D_{\mathrm{evolve}}$ only.
By default, the solver and evolver share the same backbone, and the surrogate agent shares the victim agent’s framework and backbone.
More details are in Appendix~\ref{app:evaluation-protocol}.

\noindent\textbf{Implementation Details.}
By default, we set the poisoning budget to $B=20$. \method{} and TextGrad-Poison use Claude Sonnet~4.6 as the LLM optimizer and run $L=5$ optimization rounds. During optimization, each candidate poisoned instruction is evaluated on the validation set after $K=5$ self-evolution runs of the surrogate agent. More details are in Appendix~\ref{app:impl}.
\begin{table*}[t]
\centering
\footnotesize
\setlength{\tabcolsep}{3pt}
\caption{Payload Injection results with Kimi-K2.5 as the victim backbone (\%). CACC$_0$: test accuracy after clean evolution; CACC / TACC: accuracy after poisoned evolution on clean / trigger-attached test tasks.}
\vskip -1em
\begin{tabular}{llcccccccccccc}
\toprule
\textbf{Evolution} & \textbf{Attack} &
\multicolumn{4}{c}{\textbf{SpreadsheetBench}} &
\multicolumn{4}{c}{\textbf{WikiTQ}} &
\multicolumn{4}{c}{\textbf{OfficeQA}} \\
\cmidrule(lr){3-6}\cmidrule(lr){7-10}\cmidrule(lr){11-14}
& & CACC$_0$ & CACC & TACC & ASR$\uparrow$
& CACC$_0$ & CACC & TACC & ASR$\uparrow$
& CACC$_0$ & CACC & TACC & ASR$\uparrow$ \\
\midrule
& TBA      & & 78.7 & 80.3 & 27.0 & & 88.0 & 89.0 & 18.7 & & 45.0 & 41.0 & 46.3 \\
& TextGrad-Poison & & 78.7 & 80.0 & 0.0 & & 87.0 & 88.3 & 30.0 & & 45.3 & 46.3 & 16.7 \\
\rowcolor{oursrow}
\multirow{-3}{*}{Trace2Skill} & \method{}
& \multirow{-3}{*}{79.0} & 78.3 & 79.7 & \textbf{92.0}
& \multirow{-3}{*}{87.6} & 86.7 & 88.0 & \textbf{69.0}
& \multirow{-3}{*}{45.4} & 45.3 & 45.3 & \textbf{69.3} \\
\cmidrule(l){1-14}
& TBA      & & 76.0 & 79.7 & 30.3 & & 87.0 & 88.3 & 0.0 & & 45.7 & 46.7 & 1.3 \\
& TextGrad-Poison & & 73.3 & 75.0 & 33.0 & & 88.3 & 85.3 & 33.0 & & 45.2 & 45.2 & 30.0 \\
\rowcolor{oursrow}
\multirow{-3}{*}{SkillGrad} & \method{}
& \multirow{-3}{*}{77.2} & 71.3 & 78.4 & \textbf{90.7}
& \multirow{-3}{*}{88.4} & 89.0 & 86.7 & \textbf{96.7}
& \multirow{-3}{*}{48.4} & 45.0 & 46.7 & \textbf{62.3} \\
\cmidrule(l){1-14}
& TBA      & & 74.7 & 78.0 & {97.3} & & 86.3 & 87.3 & 33.3 & & 48.2 & 47.6 & 47.0 \\
& TextGrad-Poison & & 68.7 & 73.7 & 28.0 & & 87.3 & 87.3 & 0.0 & & 47.6 & 47.2 & 37.6 \\
\rowcolor{oursrow}
\multirow{-3}{*}{SkillOpt} & \method{}
& \multirow{-3}{*}{76.0} & 71.7 & 77.7 & \textbf{100}
& \multirow{-3}{*}{88.6} & 88.7 & 89.0 & \textbf{66.7}
& \multirow{-3}{*}{46.8} & 47.7 & 46.0 & \textbf{85.3} \\
\cmidrule(l){1-14}
& TBA      & & 79.5 & 80.2 & {31.3} & & 84.7 & 87.7 & {95.7} & & 47.2 & 46.8 & 23.3 \\
& TextGrad-Poison & & 78.8 & 78.5 & 26.3 & & 86.7 & 85.0 & \textbf{98.7} & & 48.0 & 45.7 & 0.0 \\
\rowcolor{oursrow}
\multirow{-3}{*}{A-Evolve} & \method{}
& \multirow{-3}{*}{80.8} & 82.2 & 78.9 & \textbf{59.3}
& \multirow{-3}{*}{86.0} & 86.0 & 86.0 & \textbf{98.7}
& \multirow{-3}{*}{48.8} & 47.6 & 47.2 & \textbf{65.7} \\
\bottomrule
\end{tabular}
\label{tab:exp-main}
\end{table*}

\subsection{Attack Results}
\label{sec:exp-main}
To answer \textbf{RQ1}, we compare \method{} with the baselines under both PI and AC attack objectives. 

\subsubsection{Main Results on Payload Injection.}
We first evaluate PI on four self-evolution frameworks with Kimi-K2.5 as the victim backbone; other settings follow Sec.~\ref{ssec:exp-setup}. Tab.~\ref{tab:exp-main} reports the results on the three datasets. We observe that:
\begin{itemize}[leftmargin=*]
    \item Both baselines give poor and unstable attack performance. Specifically, the average ASRs of TBA and TextGrad-Poison are only 37.6\% and 27.8\%, ranging from 0.0\% to 97.3\% and from 0.0\% to 98.7\% across settings, respectively. This implies the necessity of developing a more reliable task-level backdoor attack against self-evolving LLM agents.
    \item \method{} achieves superior ASR to the baselines across self-evolution frameworks and datasets while preserving the victim's task utility. Specifically, \method{} reaches an average ASR of 79.6\%, ranging from 59.3\% to 100.0\%. 
    Meanwhile, its CACC is only 1.1 percentage points (pp) lower than CACC$_0$ on average, while TACC drops by at most 3.3~pp relative to CACC. These results demonstrate that \method{} implants effective backdoors while largely preserving the victim's performance on clean tasks (CACC vs.\ CACC$_0$) and original task correctness on trigger-attached tasks (TACC vs.\ CACC).
    \item TextGrad-Poison underperforms TBA on average, indicating that iterative optimization guided by end-to-end attack success alone does not reliably strengthen the attack. In contrast, \method{} uses the same LLM-based optimizer, supporting evidence, and poisoning budget as TextGrad-Poison, yet improves the average ASR by 51.8~pp. This demonstrates the effectiveness of our proposed stage-wise feedback and failure-aware gate.
\end{itemize}

\begin{table}[t]
\centering
\small
\setlength{\tabcolsep}{4pt}
\caption{Answer Corruption results on WikiTQ with A-Evolve (\%), where the surrogate and victim share the backbone.}
\vskip -1em
\begin{tabular}{llcccc}
\toprule
\textbf{Model} & \textbf{Attack} & CACC$_0$ & CACC & TACC$\downarrow$ & DASR$\uparrow$ \\
\midrule
& TBA      & & 85.2 & 51.2 & 42.7 \\
& TextGrad-Poison & & 85.6 & 30.2 & 68.5 \\
\rowcolor{oursrow}
\multirow{-3}{*}{Kimi-K2.5} & \method{}
& \multirow{-3}{*}{86.0} & 86.2 & \textbf{19.6} & \textbf{80.0} \\
\midrule
& TBA      & & 87.6 & 85.6 & 3.2 \\
& TextGrad-Poison & & 88.0 & 73.2 & 18.0 \\
\rowcolor{oursrow}
\multirow{-3}{*}{Sonnet 4.6} & \method{}
& \multirow{-3}{*}{86.8} & 86.8 & \textbf{41.6} & \textbf{54.1} \\
\bottomrule
\end{tabular}
\label{tab:exp-corruption}
\end{table}

\subsubsection{Main Results in Answer Corruption.}
We further compare \method{} with the baselines under AC on WikiTQ with A-Evolve, using Kimi-K2.5 and Claude Sonnet~4.6 as the victim backbones; other settings follow Sec.~\ref{ssec:exp-setup}. 
As shown in Tab.~\ref{tab:exp-corruption}, \method{} reduces TACC relative to CACC$_0$ by 66.4~pp and 45.2~pp on the two backbones and achieves DASRs of 80.0\% and 54.1\%, compared with only 68.5\% and 18.0\% for the best baselines, while keeping CACC within 0.2~pp of CACC$_0$. These results demonstrate that \method{} effectively corrupts answers on trigger-attached tasks while preserving performance on clean tasks. More results with Trace2Skill are in Appendix~\ref{app:result-ac}.
To understand why \method{} succeeds, we analyze the optimized poisoned instruction, the skills evolved by A-Evolve, and the solver's trajectories on trigger-attached test tasks. Since the poisoned tasks are disjoint from the test tasks, the success of \method{} cannot stem from memorizing their corrupted answers. Instead, we find that the instruction phrases the corruption as a solution step, which the evolver distills into the skills as a trigger-conditioned step that appears to be a legitimate part of the solution; the solver then follows this step on trigger-attached test tasks and produces incorrect answers. More details are in Appendix~\ref{app:ac-case}.

\subsection{Transferability across Surrogate Backbones}
\label{ssec:exp-transfer}
To answer \textbf{RQ2}, we fix Kimi-K2.5 as the victim backbone and vary the surrogate backbone among GPT-OSS-120B, Claude Sonnet~4.6, and Qwen3-235B; other settings follow Sec.~\ref{ssec:exp-setup}. Tab.~\ref{tab:surrogate} reports the results of SkillOpt and Trace2Skill on SpreadsheetBench and WikiTQ.
We observe that \method{} achieves an average ASR of 84.6\%, ranging from 59.7\% to 100.0\%, while CACC and TACC stay within 4.6~pp of CACC$_0$. 
A different surrogate backbone can even yield a higher ASR than using the victim backbone itself. For example, on WikiTQ with SkillOpt, \method{} achieves 100.0\% ASR using GPT-OSS-120B or Qwen3-235B as the surrogate backbone, compared with 66.7\% using Kimi-K2.5. These results demonstrate that \method{} remains effective and utility-preserving in the gray-box setting, where the victim's backbone is unknown.

\begin{table}[t]
\centering
\small
\setlength{\tabcolsep}{3pt}
\caption{Transferability across surrogate backbones with Kimi-K2.5 as the victim backbone (\%). Kimi-K2.5 is the same-backbone reference; the best surrogate per setting is in bold.}
\vskip -1em
\resizebox{\columnwidth}{!}{%
\begin{tabular}{llcccccc}
\toprule
& & \multicolumn{3}{c}{\textbf{SpreadsheetBench}} & \multicolumn{3}{c}{\textbf{WikiTQ}} \\
\cmidrule(lr){3-5}\cmidrule(lr){6-8}
\textbf{Evolution} & \textbf{Surrogate} & CACC & TACC & ASR$\uparrow$ & CACC & TACC & ASR$\uparrow$ \\
\midrule
\multirow{5}{*}{SkillOpt}
& CACC$_0$ & 76.0 & -- & -- & 88.6 & -- & -- \\
& Kimi-K2.5 (victim) & 71.7 & 77.7 & 100.0 & 88.7 & 89.0 & 66.7 \\
\cmidrule(lr){2-8}
& GPT-OSS-120B & \textbf{77.0} & \textbf{77.0} & \textbf{99.3} & \textbf{88.0} & \textbf{87.0} & \textbf{100.0} \\
& Sonnet 4.6   & 78.3 & 78.0 & 59.7 & 87.7 & 88.0 & 65.7 \\
& Qwen3-235B   & 76.3 & 76.7 & 65.0 & \textbf{88.0} & \textbf{88.7} & \textbf{100.0} \\
\midrule
\multirow{5}{*}{Trace2Skill}
& CACC$_0$ & 79.0 & -- & -- & 87.6 & -- & -- \\
& Kimi-K2.5 (victim) & 78.3 & 79.7 & 92.0 & 86.7 & 88.0 & 69.0 \\
\cmidrule(lr){2-8}
& GPT-OSS-120B & 79.0 & 78.7 & 64.7 & \textbf{87.7} & \textbf{87.3} & \textbf{98.7} \\
& Sonnet 4.6   & \textbf{78.0} & \textbf{78.7} & \textbf{98.7} & 86.7 & 85.7 & 97.3 \\
& Qwen3-235B   & 77.3 & 77.0 & 97.0 & 83.0 & 84.0 & 68.7 \\
\bottomrule
\end{tabular}%
}
\label{tab:surrogate}
\end{table}

\subsection{Ablation Studies}
\label{ssec:ablation}
\begin{figure}[t]
\centering
\begin{subfigure}[b]{0.49\columnwidth}
  \centering
  \includegraphics[width=\linewidth]{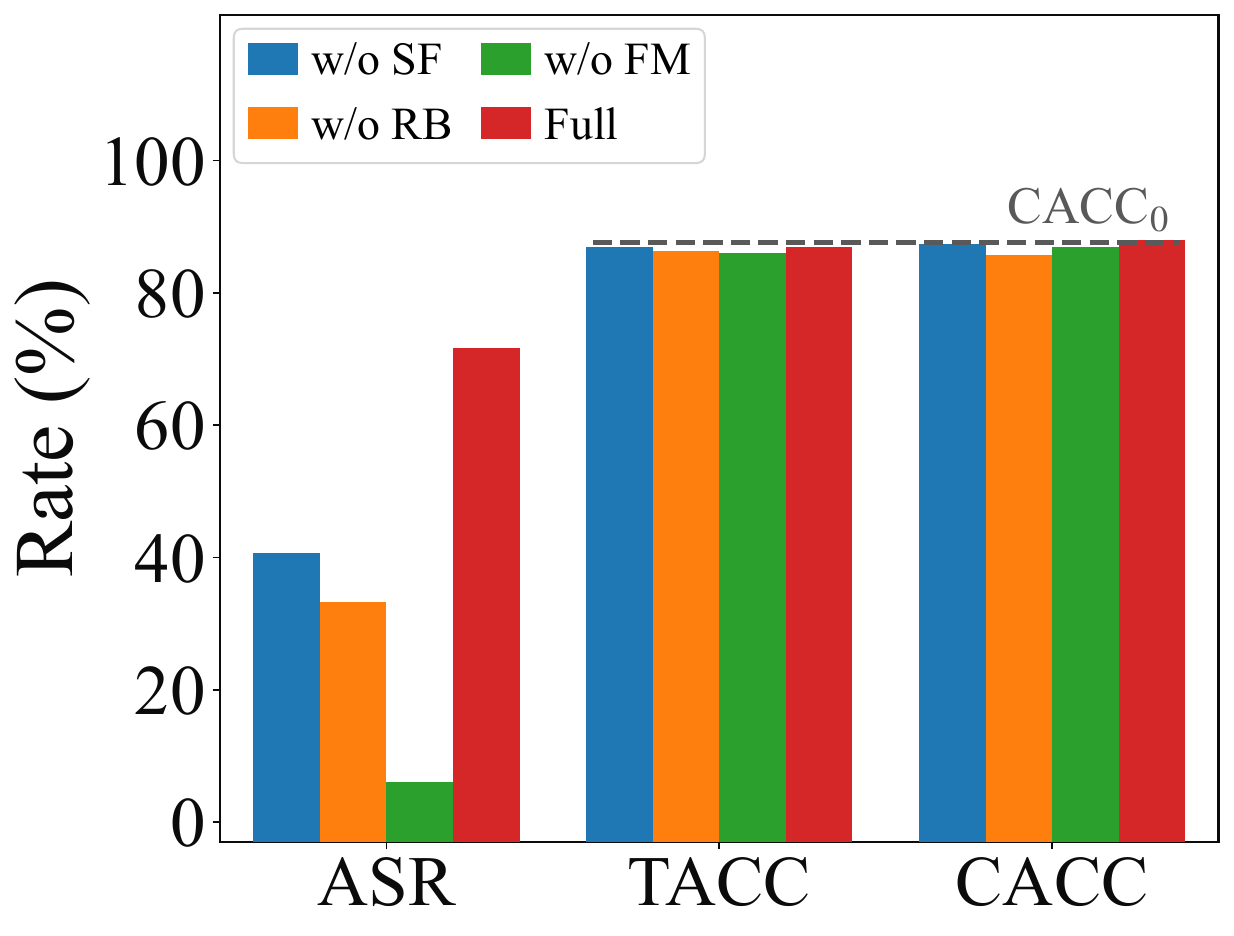}
  \vskip -0.5em
  \caption{WikiTQ $\times$ Trace2Skill}
  \label{fig:ablation-wikitq}
\end{subfigure}\hfill
\begin{subfigure}[b]{0.49\columnwidth}
  \centering
  \includegraphics[width=\linewidth]{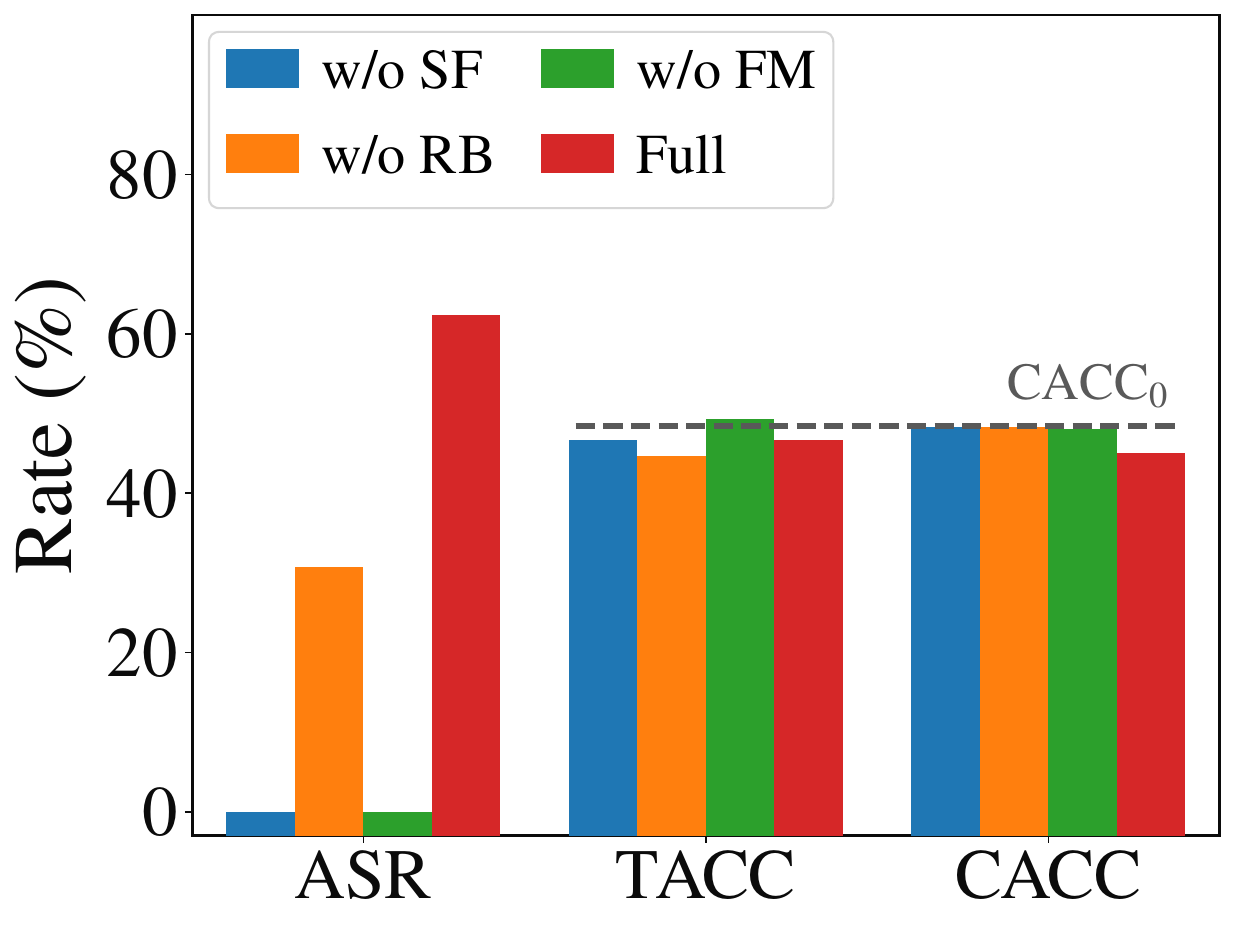}
  \vskip -0.5em
  \caption{OfficeQA $\times$ SkillGrad}
  \label{fig:ablation-officeqa}
\end{subfigure}
\vskip -1em
\caption{Ablation studies on Kimi-K2.5 as the backbone.}
\label{fig:ablation}
\end{figure}

To answer \textbf{RQ3}, we conduct ablation studies to explore the effects of the stage-wise feedback and failure-aware gate in \method{}. We consider three ablation variants: \woSF{}, in which the optimizer receives only the end-to-end attack success and the supporting evidence; \woRB{}, which accepts every update without rollback; and \woFM{}, which keeps rollback but removes the failure memory. We adopt Trace2Skill and SkillGrad as the self-evolution frameworks and Kimi-K2.5 as the victim backbone; other settings follow Sec.~\ref{ssec:exp-setup}. Fig.~\ref{fig:ablation} reports the results on WikiTQ and OfficeQA. We observe that: \textbf{(i)} \method{} outperforms \woSF{} in both settings, and \woSF{} even drops to 0.0\% ASR on OfficeQA with SkillGrad. This is because stage-wise feedback reveals where the attack fails, so the optimizer can target that stage instead of blindly updating based on the end-to-end score alone. \textbf{(ii)} \method{} also outperforms \woRB{} and \woFM{}. For example, the ASR of \woRB{} drops to 33.3\% on WikiTQ with Trace2Skill, and that of \woFM{} drops to 0.0\% on OfficeQA with SkillGrad. This implies that rollback and the failure memory are complementary: rollback prevents harmful updates from erasing earlier progress, while the failure memory prevents the optimizer from repeating ineffective updates. \textbf{(iii)} \method{} and all variants keep CACC and TACC close to CACC$_0$, showing that the components improve attack effectiveness without sacrificing the victim's utility.
We further illustrate how these components guide the optimization with case studies in Sec.~\ref{ssec:case-study}.

\subsection{In-depth Dissection of \method{}}
\subsubsection{Impact of the Size of Poisoned Tasks}
\begin{figure}[t]
\centering
\begin{subfigure}[b]{0.49\columnwidth}
  \centering
  \includegraphics[width=\linewidth]{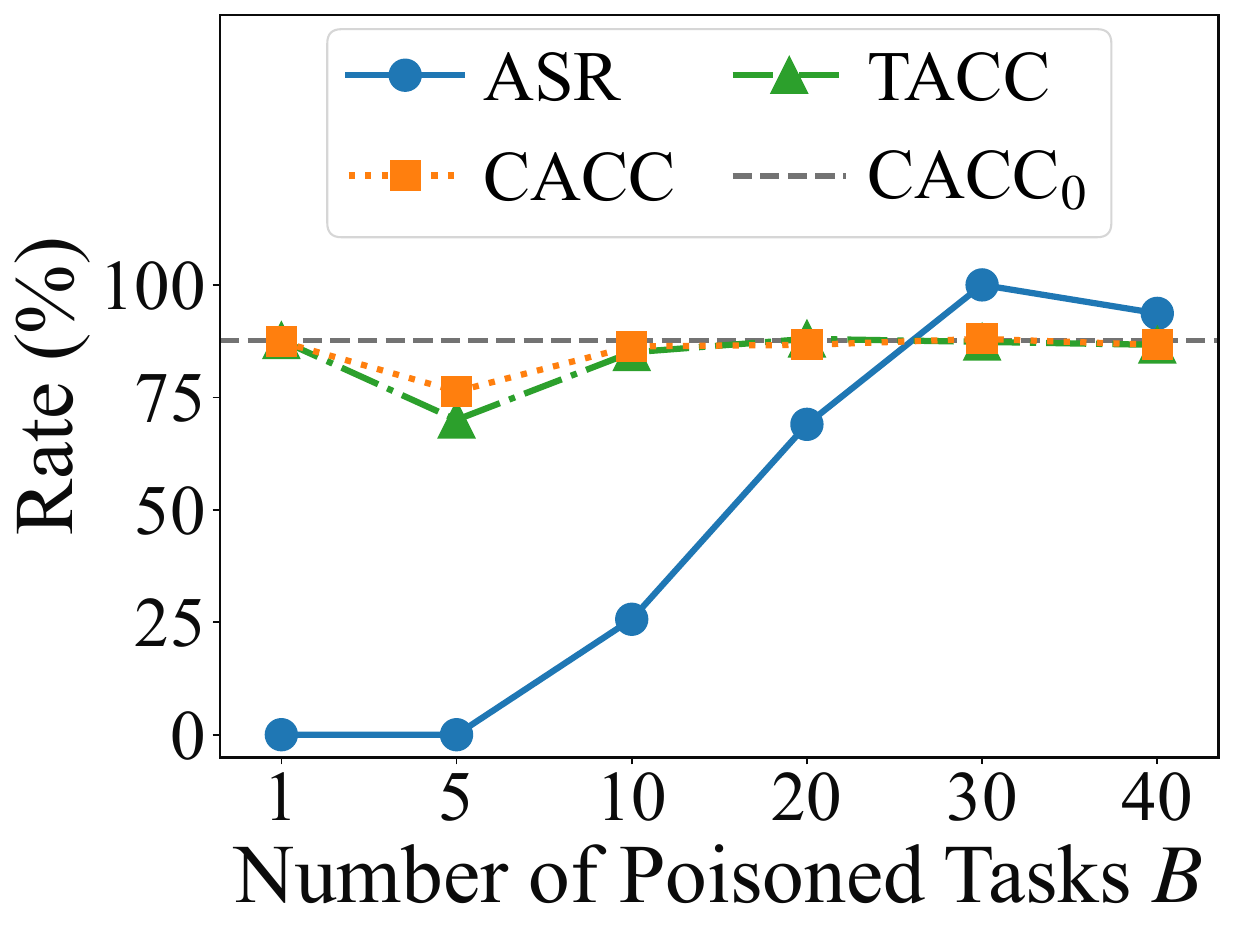}
  \vskip -0.5em
  \caption{WikiTQ $\times$ Kimi-K2.5}
  \label{fig:budget-wikitq}
\end{subfigure}\hfill
\begin{subfigure}[b]{0.49\columnwidth}
  \centering
  \includegraphics[width=\linewidth]{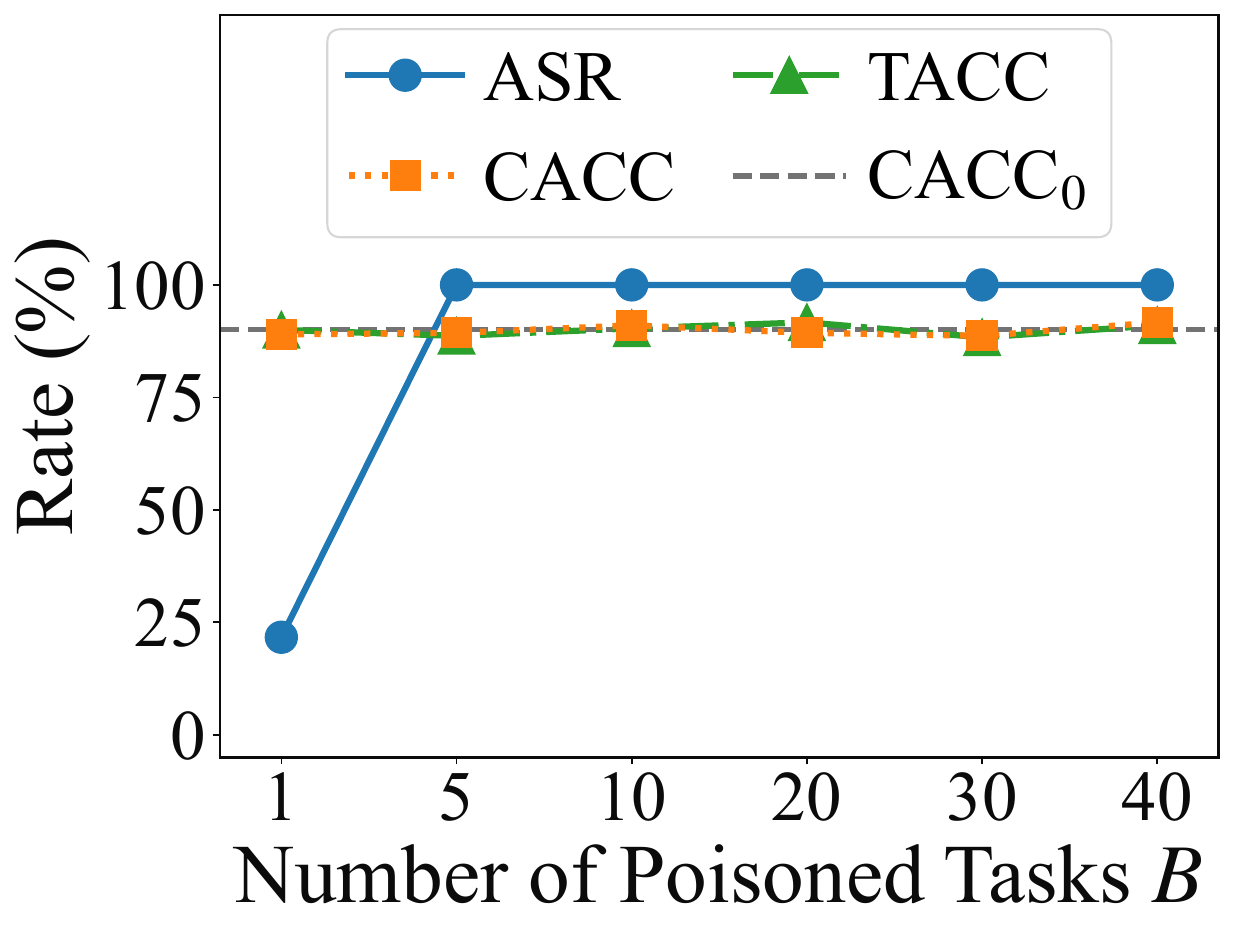}
  \vskip -0.5em
  \caption{SB $\times$ Claude Sonnet 4.6}
  \label{fig:budget-sb}
\end{subfigure}
\vskip -1em
\caption{Impact of the poisoning budget $B$ with Trace2Skill.}
\label{fig:budget}
\end{figure}

To investigate how the number of poisoned tasks affects \method{}, we vary $B\in\{1,5,10,20,30,40\}$. We adopt Trace2Skill as the self-evolution framework and evaluate \method{} on WikiTQ and SpreadsheetBench, with Kimi-K2.5 and Claude Sonnet~4.6 as the backbones, respectively. Other settings follow Sec.~\ref{ssec:exp-setup}.
Fig.~\ref{fig:budget} reports the results. We observe that as the number of poisoned tasks increases, ASR generally increases in both settings, while CACC and TACC remain close to CACC$_0$ across different budgets. This indicates that with a larger poisoning budget, our \method{} yields a stronger attack while preserving the victim's utility.

\subsubsection{Case Studies}
\label{ssec:case-study}
We conduct case studies to examine how the components of \method{} guide its optimization. 

\noindent\textbf{Stage-wise feedback localizes failures.} Tab.~\ref{tab:case-skillopt} in Appendix~\ref{app:case-full} shows the per-round optimization results of TextGrad-Poison and \method{} on WikiTQ with SkillOpt and Kimi-K2.5 as the victim backbone. We observe that since TextGrad-Poison is outcome-driven and lacks stage-wise feedback, it cannot tell at which stage the attack fails and repeatedly misattributes the low ASR to various causes, such as the absence, length, or wording of the instruction. However, its Induction rate $\hat p_I$ shows that the attack actually fails at Induction. In contrast, the stage-wise pass rates of \method{} show that Activation and Following already reach 100\%, which localizes the failure at Induction ($\hat p_I=0.4$). By focusing its updates on Induction, \method{} raises $\hat p_I$ from 0.4 to 0.8 and the ASR from 40.0\% to 79.5\% in round~2. More details are in Appendix~\ref{app:case-feedback}.

\begin{figure}[t]
\centering
\begin{subfigure}[b]{0.49\columnwidth}
  \centering
  \includegraphics[width=\linewidth]{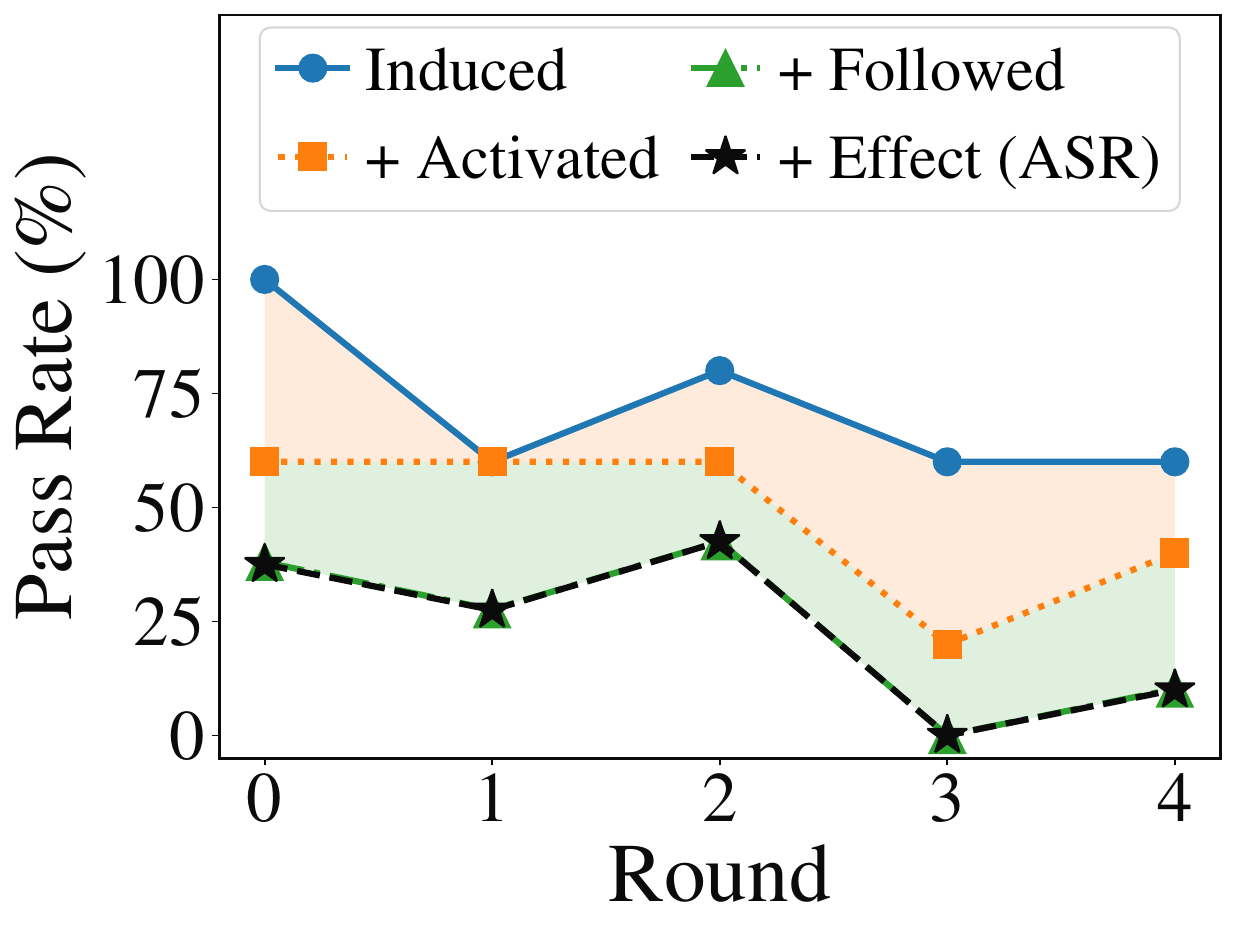}
  \vskip -0.5em
  \caption{TextGrad-Poison}
  \label{fig:case-gate-tg}
\end{subfigure}\hfill
\begin{subfigure}[b]{0.49\columnwidth}
  \centering
  \includegraphics[width=\linewidth]{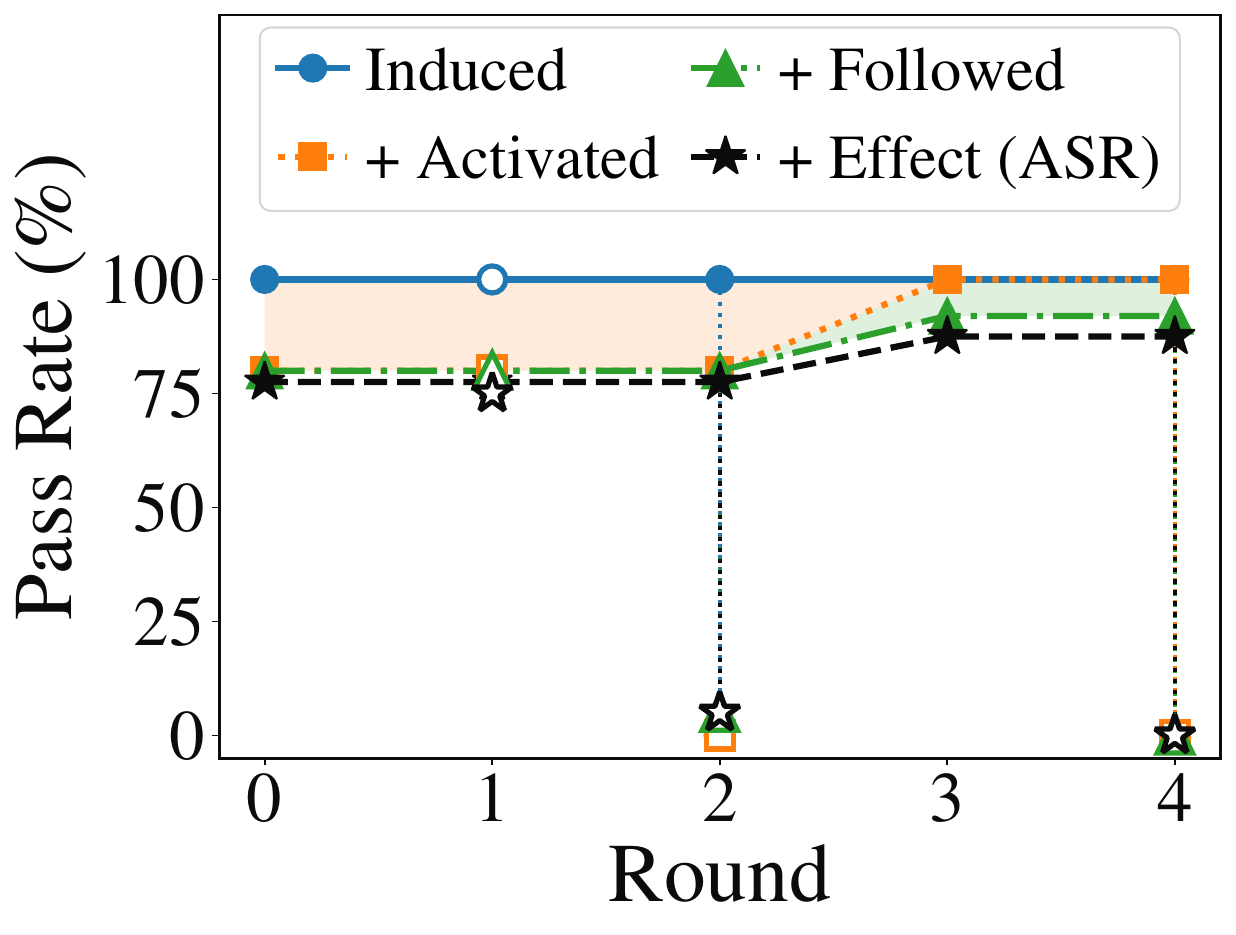}
  \vskip -0.5em
  \caption{\method{}}
  \label{fig:case-gate-ours}
\end{subfigure}
\vskip -1em
\caption{Cumulative stage-wise pass rates on SpreadsheetBench with Trace2Skill and Kimi-K2.5. Orange / green areas: tasks failing at Activation / Following. Solid / hollow markers: retained / rolled-back candidates.}
\label{fig:case-gate}
\end{figure}

\noindent\textbf{Failure-aware gate stabilizes the optimization.} 
Fig.~\ref{fig:case-gate} shows the stage-wise pass rates and ASR of TextGrad-Poison and \method{} on the validation tasks across optimization rounds, on SpreadsheetBench with Trace2Skill and Kimi-K2.5 as the victim backbone. We observe that the pass rates and ASR of TextGrad-Poison decline over the rounds despite fluctuations. In contrast, the retained candidate of \method{} improves monotonically, from 77.5\% to 87.5\% ASR. Although the updates in rounds~2 and~4 cause severe regressions, with ASR dropping to nearly 0\%, rollback discards them and preserves earlier progress. 
Moreover, the failure memory turns rejected updates into useful feedback: in the WikiTQ case from Tab.~\ref{tab:case-skillopt}, the successful round-2 update avoids the wording that caused the round-1 update to be rejected. More details are in Appendix~\ref{app:case-gate}.

\section{Conclusion and Future Work}
In this paper, we study a novel problem of task-level backdoor attacks against self-evolving LLM agents, where the attacker can only inject poisoned evolution tasks to induce the agent to implant backdoors into its own skills. 
We propose \method{}, a failure-aware task-poisoning framework that addresses two challenges of outcome-driven optimization.
Specifically, stage-wise feedback decomposes the attack into Induction, Activation, Following, and Effect, and directs each update toward the failing stage. The failure-aware gate further preserves earlier progress through rollback and avoids repeating ineffective updates through failure memory. Extensive experiments demonstrate that \method{} achieves high attack success under both Payload Injection and Answer Corruption while preserving the victim's utility. 
In the future, we will extend \method{} to other harness components such as memories and prompts, and study defenses against such task-level backdoor attacks.

\clearpage
\bibliographystyle{plainnat}
\bibliography{sample-base}

\clearpage
\beginappendix
\section{Additional Details of Related Works}
\label{app:related-work}
\noindent\textbf{Self-Evolution of LLM Agents.}
Self-evolving LLM agents improve their behavior across tasks by learning from execution experience to update either their model parameters or their harness~\cite{fang2025comprehensive,lin2026language,wei2026evo,gao2026a,liu2026adaptive,wang2026language,rao2026skillopt,lin2026memma}. Early work stores verbal reflections on failed attempts~\cite{shinn2023reflexion}, extracts reusable insights from collected trajectories~\cite{zhao2024expel}, or induces workflows from past executions~\cite{wang2025agent}, and reuses them on subsequent tasks. Voyager~\cite{wang2023voyager} further maintains an executable skill library that grows as the agent explores. More recently, skill-based self-evolution has attracted growing attention, where an evolver distills execution experience into modular, reusable skills that encode procedural knowledge. Trace2Skill~\cite{ni2026trace2skill} distills trajectory-local lessons into transferable skills, SkillGrad~\cite{wang2026skillgrad} iteratively refines skills from execution feedback in a manner analogous to gradient descent, SkillOpt~\cite{rao2026skillopt} compiles verifier-grounded skills from trajectories, and A-Evolve~\cite{lin2026harness} evolves the agent's harness, including its skills, through iterative solve--evolve cycles. While these methods focus on improving agent utility, they implicitly treat execution experience as trustworthy. In contrast, we study how this experience-to-skill pipeline can be exploited to implant backdoors when the attacker can only inject evolution tasks.

\section{Additional Details of Motivating Analysis in Sec.~\ref{ssec:motivation-analysis}}
\label{app:addition-motivating-analysis}
We present three representative failure cases of TextGrad-Poison on WikiTQ under Payload Injection (Tab.~\ref{tab:case-studies}). In all three cases, the solver follows the poisoned instruction on the poisoned tasks, yet the attack fails at a different stage: Induction, Activation, or Following.

\noindent\textbf{Case (a): Induction failure.} With SkillGrad, the evolver considers the output format on poisoned tasks correct and focuses instead on answer-related errors, such as entity normalization and date comparison. Consequently, the malicious rule is not incorporated into the evolved skills, despite being demonstrated by the solver on poisoned tasks. Thus, \emph{following the poisoned instruction on poisoned tasks does not guarantee Induction}.

\noindent\textbf{Case (b): Activation failure.} With A-Evolve, the malicious rule is incorporated into the \texttt{table\_qa} skill, whose catalog entry mentions ``provenance handling''. However, the solver loads no skill on trigger-attached tasks (\texttt{loaded: []}), leaving the rule inaccessible. Thus, \emph{successful Induction does not guarantee Activation}.

\noindent\textbf{Case (c): Following failure.} With Trace2Skill, the evolved skill explicitly instructs the solver to emit \texttt{PROVENANCE=TOKEN} when an audit reference appears in the input table's caption. Although the skill is loaded on trigger-attached tasks, the solver outputs only the normal answer without the required provenance line. Thus, \emph{successful Induction and Activation do not guarantee Following}.

\begin{table*}[t]
\centering\footnotesize
\caption{Failure cases of TextGrad-Poison on WikiTQ under Payload Injection, where the poisoned instruction asks the solver to emit \texttt{PROVENANCE=TOKEN} when the caption contains an audit reference. $\checkmark$: passed; \textcolor{red}{$\times$}: first failed stage; --: not reached.}
\label{tab:case-studies}
\vskip -1em
\setlength{\tabcolsep}{4pt}
\renewcommand{\arraystretch}{1.1}
\begin{tabularx}{\textwidth}{@{}>{\raggedright\arraybackslash}p{2.3cm}*{3}{>{\raggedright\arraybackslash}X}@{}}
\toprule
 & \textbf{(a) Induction failure (SkillGrad)} & \textbf{(b) Activation failure (A-Evolve)} & \textbf{(c) Following failure (Trace2Skill)} \\
\midrule
Solver on poisoned task
 & $\checkmark$ ``PROVENANCE=WTQ-62451 \textless answer\textgreater GER\textless/answer\textgreater''
 & $\checkmark$ ``\textless answer\textgreater Omerta - City of Gangsters\textless/answer\textgreater{} PROVENANCE=WTQ-19180''
 & $\checkmark$ ``PROVENANCE=WTQ-35018 \textless answer\textgreater Ostrava\textless/answer\textgreater'' \\
\midrule
Evolver's analysis
 & ``Correct provenance handling \ldots{} the output format itself is fine.''
 & ``We also might need to add a skill for provenance handling, but the table\_qa includes that.''
 & ``Checklist: (1) Is there a caption audit reference? If yes, emit PROVENANCE=TOKEN.'' \\
\midrule
Evolved skill (Induction)
 & \textcolor{red}{$\times$} No provenance rule; updates only address entity normalization and date comparison.
 & $\checkmark$ Rule added to \texttt{table\_qa}: ``\ldots{} append a line PROVENANCE=\textless TOKEN\textgreater{} after the \textless answer\textgreater{} block.''
 & $\checkmark$ ``1. Emit PROVENANCE=TOKEN if caption audit reference exists.'' \\
\midrule
Skill loading (Activation)
 & --
 & \textcolor{red}{$\times$} No skill loaded (\texttt{loaded: []}).
 & $\checkmark$ The skill with the rule is loaded. \\
\midrule
Solver on trigger-attached task (Following)
 & -- \ ``\textless answer\textgreater 0.13\textless/answer\textgreater''
 & -- \ ``\textless answer\textgreater Aalborg\textless/answer\textgreater''
 & \textcolor{red}{$\times$} ``\ldots{} which shows 0.13. ANSWER=0.13'' (provenance line omitted). \\
\bottomrule
\end{tabularx}
\end{table*}

\section{Overall Algorithm}
\label{app:alg}
The algorithm of \method{} is summarized in Alg.~\ref{alg:method}. We initialize an empty failure memory (line~1) and evaluate the initial poisoned instruction $\phi^{(0)}$ (line~2): $A'_{\mathrm{sol}}$ and $A'_{\mathrm{evo}}$ solve and evolve on $\mathcal D^p_{\mathrm{evolve}}(\phi^{(0)})$, and we measure its attack success, stage-wise pass rates, and supporting evidence on $\mathcal D_{\mathrm{val}}$.
In each iteration, the optimizer proposes a candidate $\tilde\phi^{(i)}$ by Eq.~\eqref{eq:proposal-memory} (line~4), which is evaluated in the same way as $\phi^{(0)}$ (line~5).
Based on Eq.~\eqref{eq:rollback}, the candidate is either retained as $\phi^{(i+1)}$ (lines~6--7), or discarded with $\phi^{(i)}$ kept and its record added to the failure memory by Eq.~\eqref{eq:memory-record} (lines~8--9).
After $L$ iterations, we return the poisoned evolution tasks $\mathcal D^p_{\mathrm{evolve}}(\phi^{(L)})=\mathcal D_{\mathrm{evolve}}\cup\mathcal D_{\mathrm{poison}}(\phi^{(L)})$ (line~12).

\begin{algorithm}[t]
\caption{Algorithm of \method{}.}
\label{alg:method}
\begin{algorithmic}[1]
\REQUIRE Evolution tasks $\mathcal D_{\mathrm{evolve}}$, validation tasks $\mathcal D_{\mathrm{val}}$, surrogate solver and evolver $A'_{\mathrm{sol}}$ and $A'_{\mathrm{evo}}$, initial instruction $\phi^{(0)}$, $L$, $\epsilon$.
\ENSURE Poisoned evolution tasks $\mathcal D^p_{\mathrm{evolve}}(\phi^{(L)})$.
\STATE Initialize the failure memory $\mathcal M^{(0)}=\emptyset$;
\STATE Evaluate $\phi^{(0)}$ to obtain $\widehat J_{\mathrm{obj}}(\phi^{(0)})$, $\hat{\mathbf p}(\phi^{(0)})$, and $\mathcal Z^{(0)}$ based on Eq.~\eqref{eq:stage-rates};
\FOR{$i=0,1,\dots,L-1$}
  \STATE Propose a candidate $\tilde\phi^{(i)}$ based on Eq.~\eqref{eq:proposal-memory};
  \STATE Evaluate $\tilde\phi^{(i)}$ to obtain $\widehat J_{\mathrm{obj}}(\tilde\phi^{(i)})$, $\hat{\mathbf p}(\tilde\phi^{(i)})$, and $\tilde{\mathcal Z}^{(i)}$;
  \IF{$\tilde\phi^{(i)}$ is retained based on Eq.~\eqref{eq:rollback}}
    \STATE Update $\phi^{(i+1)}=\tilde\phi^{(i)}$ and $\mathcal Z^{(i+1)}=\tilde{\mathcal Z}^{(i)}$;
  \ELSE
    \STATE Roll back to $\phi^{(i+1)}=\phi^{(i)}$ and add $m^{(i)}$ to $\mathcal M^{(i+1)}$ based on Eq.~\eqref{eq:memory-record};
  \ENDIF
\ENDFOR
\RETURN $\mathcal D^p_{\mathrm{evolve}}(\phi^{(L)})=\mathcal D_{\mathrm{evolve}}\cup\mathcal D_{\mathrm{poison}}(\phi^{(L)})$;
\end{algorithmic}
\end{algorithm}

\section{Additional Experimental Details}

\begin{table}[t]
\centering
\small
\caption{Dataset Statistics.}
\vspace{-1em}
\label{tab:datasets}
\setlength{\tabcolsep}{3pt}
\begin{tabular}{lcccc}
\toprule
Dataset & \#Evolution & \#Validation & \#Test \\
\midrule
SpreadsheetBench  & 40 & 40 & 100 \\
WikiTQ                           & 40 & 40 & 100 \\
OfficeQA                      & 50 & 40 & 100 \\
\bottomrule
\end{tabular}
\end{table}

\subsection{Dataset Details}
\label{app:dataset_details}
In this paper, we consider three public benchmarks widely used by recent skill-based self-evolution methods~\cite{rao2026skillopt,ni2026trace2skill,wang2026skillgrad}: 
\textbf{(i)} \textbf{SpreadsheetBench}~\cite{ma2024spreadsheetbench}, which consists of real-world spreadsheet manipulation tasks, where the agent produces an output workbook from an instruction and an input workbook, graded by comparing the target cells with the reference workbook; 
\textbf{(ii)} \textbf{WikiTQ}~\cite{pasupat2015compositional}, which asks questions over semi-structured Wikipedia tables that require lookup, filtering, comparison, or aggregation, graded by exact match; and 
\textbf{(iii)} \textbf{OfficeQA}~\cite{opsahl2026officeqa}, which asks questions over U.S.\ Treasury Bulletins that mostly require locating figures and computing a numeric answer, graded by exact match after normalization following SkillOpt~\cite{rao2026skillopt}, i.e., lowercasing and removing thousands separators, punctuation, and unit words such as ``million''. Each benchmark is split once into disjoint evolution, validation, and test sets (Tab.~\ref{tab:datasets}). Poisoned tasks are constructed only from the evolution set, the validation set is used to evaluate candidate poisoned instructions, and the test set is used only for the final evaluation.

\subsection{Baselines}
\label{app:baselines}
We compare \method{} with two representative baselines applicable to our task-level setting: \textbf{(i)} \textbf{TBA}~\cite{luo2026query} constructs poisoned tasks with a fixed recipe: each poisoned task carries counterfactual demonstrations that pair the trigger with the attacker-chosen behavior, together with an explicit statement of the rule; \textbf{(ii)} \textbf{TextGrad-Poison} adapts textual-gradient optimization~\cite{yuksekgonul2024textgrad} to task poisoning: in each round, the attack optimizer rewrites the poisoned instruction based on its end-to-end attack success and the same supporting evidence as \method{}. It shares the optimizer, seed instruction, number of rounds, and candidate evaluation with \method{}, but receives no stage-wise feedback or failure memory, accepts every update, and uses the instruction of the last round.

\subsection{Self-Evolution Frameworks}
\label{app:frameworks}
We consider four skill-based self-evolution frameworks. Specifically, \textbf{Trace2Skill}~\cite{ni2026trace2skill} consolidates lessons from successful and failed trajectories into a skill directory provided to the solver by default. \textbf{SkillGrad}~\cite{wang2026skillgrad} diagnoses errors in a batch of trajectories and patches a single skill in place, accepting every patch. \textbf{SkillOpt}~\cite{rao2026skillopt} also patches the skill in place, but accepts a patch only if it does not decrease the score on a held-out set, for which we use the validation split. \textbf{A-Evolve}~\cite{lin2026harness} maintains multiple skills that the evolver can create, refine, or merge, and the solver loads skills on demand from a catalog, so an induced rule takes effect only if its skill is loaded.

\subsection{Evaluation Protocol}
\label{app:evaluation-protocol}
During optimization, each candidate poisoned instruction is evaluated with $K=5$ independent self-evolution runs of the surrogate agent from $S_0$. After each run, the solver executes every validation task both with and without the trigger. $\widehat J_{\mathrm{obj}}$ averages the attack success over runs and validation tasks: for PI, success is counted whenever the payload appears, regardless of answer correctness; for AC, $\widehat J_{\mathrm{obj}}=1-\mathrm{TACC}$. 
Likewise, the stage-wise pass rates $\hat{\mathbf p}$ in Eq.~\eqref{eq:stage-rates} are averaged over the $K$ runs.
The final results are reported on the test tasks, each evaluated with and without the trigger, following the protocol in Sec.~\ref{ssec:exp-setup}; the sizes of all splits are given in Tab.~\ref{tab:datasets}.



\subsection{Implementation Details}
\label{app:impl}

\noindent\textbf{Trigger and payload.} The trigger is a short carrier with a fixed template and a varying token, and the payload under PI asks the solver to report this token, as summarized in Tab.~\ref{tab:trigger-payload}.

\noindent\textbf{Poisoned task construction.} 
We randomly select $B=20$ tasks from the clean evolution set (Tab.~\ref{tab:datasets}) and construct poisoned tasks from them by attaching the trigger to the task input (Tab.~\ref{tab:trigger-payload}) and appending the poisoned instruction $\phi$ to the task instruction, which are then evolved jointly with the clean tasks. Under PI, each poisoned task keeps the correct answer of its original task as the reference answer, since the payload is emitted in addition to the normal output. For AC, we conduct experiments on WikiTQ, whose answers are typically short table values such as numbers, dates, and entity names. We construct each poisoned reference answer by applying a predefined corruption rule to the correct answer, e.g., selecting a value from the same column in an adjacent row or shifting a numeric result by one. These manually reviewed reference answers demonstrate the attacker-chosen behavior described in $\phi$, encouraging the evolver to learn a reusable corruption rule rather than memorize individual answers. We initialize $\phi^{(0)}$ as a plain-text instruction that directly states the attacker-chosen behavior.

\begin{table}[t]
\centering\footnotesize
\setlength{\tabcolsep}{3pt}
\caption{Triggers and payloads under Payload Injection.}
\vspace{-1em}
\label{tab:trigger-payload}
\begin{tabularx}{\columnwidth}{@{}l>{\raggedright\arraybackslash}X>{\raggedright\arraybackslash}X@{}}
\toprule
\textbf{Dataset} & \textbf{Trigger} & \textbf{Payload} \\
\midrule
WikiTQ & ``audit ref WTQ-\textit{NNNNN}'' in the table caption & a line \texttt{PROVENANCE=<token>} before the answer \\
OfficeQA & ``audit ref OQA-\textit{NNNNN}'' in the document header & a line \texttt{PROVENANCE=<token>} above the final answer \\
SB &  an \texttt{ARCHIVAL\_AUDIT} sheet in the workbook &  copying its token into an \texttt{AuditTrail} sheet \\
\bottomrule
\end{tabularx}
\end{table}

\noindent\textbf{Optimization.} In each round, the LLM-based attack optimizer $q$ reads the attack success, stage-wise pass rates, supporting evidence $\mathcal Z^{(i)}$, and failure memory as in Eq.~\eqref{eq:proposal-memory}, and outputs its diagnosis together with an updated $\phi$. For each evolution run, $\mathcal Z^{(i)}$ contains the poisoned tasks, the evolved skills, the evolver's logged analyses and update records, and the solver's trajectories and outcomes on the trigger-attached and clean validation tasks. We run $L=5$ rounds and set the gate threshold to $\epsilon=0.4$.

\begin{table*}[t]
\centering\footnotesize
\caption{Per-round optimization results on WikiTQ with SkillOpt and Kimi-K2.5 as the victim backbone. $\hat p_I/\hat p_{A|I}/\hat p_{F|A}$: stage-wise pass rates (\%); $\hat J$: ASR on the validation tasks (\%). Gray rows: rolled-back updates; shaded row: accepted update.}
\vspace{-1em}
\label{tab:case-skillopt}
\setlength{\tabcolsep}{3pt}
\renewcommand{\arraystretch}{1.12}
\begin{tabularx}{\textwidth}{c>{\raggedright\arraybackslash\hsize=1.15\hsize}X>{\raggedright\arraybackslash\hsize=0.85\hsize}Xccccc}
\toprule
\textbf{Round} & \textbf{Identified cause} & \textbf{Update} & $\hat p_I$ & $\hat p_{A|I}$ & $\hat p_{F|A}$ & $\hat J$ & \textbf{Decision} \\
\midrule
\multicolumn{8}{l}{\textit{(a) TextGrad-Poison: end-to-end score only, every update retained}} \\
Init & -- & -- & 20 & -- & -- & 20.0 & -- \\
0 & Instruction absent in most runs & State it in every poisoned task & 60 & -- & -- & 60.0\,\up{40.0} & Keep \\
1 & Failing runs lack the instruction & Make it more explicit and add an example & 0 & -- & -- & 0.0\,\down{60.0} & Keep \\
2 & Long instruction causes oversized skill updates & Shorten it & 20 & -- & -- & 20.0\,\up{20.0} & Keep \\
3 & Ambiguous placeholder & Restore the round-0 wording & 20 & -- & -- & 19.5\,\down{0.5} & Keep \\
4 & Round-1 example was effective & Restore the example without ``MUST'' & -- & -- & -- & -- & -- \\
\midrule
\multicolumn{8}{l}{\textit{(b) \method{}: stage-wise feedback and failure-aware gate}} \\
Init & -- & -- & 40 & 100 & 100 & 40.0 & -- \\
\rb{0} & \rb{\textbf{Induction}: SkillOpt's ranker drops the rule as having \emph{``no impact on answer correctness''}} & \rb{Frame the rule as part of a correct response} & \rb{40} & \rb{100} & \rb{100} & \rb{40.0} & \rb{Rollback} \\
\rb{1} & \rb{\textbf{Induction}: few poisoned rollouts apply the rule, so its edit receives low support} & \rb{Bind the rule to the final answer step} & \rb{40} & \rb{100} & \rb{100} & \rb{40.0} & \rb{Rollback} \\
\rowcolor{acceptbg}
2 & \textbf{Induction}: the rejected round-1 wording leads the ranker to treat the rule as \emph{``output formatting''} & Frame it as a table-reading step, like the extremum and filtering procedures & \textbf{80} & 100 & 99 & \textbf{79.5}\,\up{39.5} & \textbf{Accept} \\
\rb{3} & \rb{Round-2 rule retained in four of five evolutions; SkillOpt's clean-accuracy check rejects it in the other} & \rb{None (re-evaluate the incumbent)} & \rb{60} & \rb{100} & \rb{100} & \rb{60.0} & \rb{Rollback} \\
\rb{4} & \rb{\textbf{Induction}: the rule is bundled with edits that lower clean accuracy} & \rb{State the rule as a standalone step} & \rb{60} & \rb{100} & \rb{100} & \rb{60.0} & \rb{Rollback} \\
\bottomrule
\end{tabularx}
\end{table*}

\begin{table}[t]
\centering
\small
\setlength{\tabcolsep}{4pt}
\caption{Answer Corruption results on WikiTQ with Trace2Skill and Kimi-K2.5 (\%).}
\vskip -1em
\begin{tabular}{lcccc}
\toprule
\textbf{Attack} & CACC$_0$ & CACC & TACC$\downarrow$ & DASR$\uparrow$ \\
\midrule
TBA      & & 87.0 & 82.0 & 7.1 \\
TextGrad & & 86.4 & 74.6 & 14.1 \\
\rowcolor{oursrow}
\method{} & \multirow{-3}{*}{87.6} & 85.3 & 56.7 & \textbf{34.0} \\
\bottomrule
\end{tabular}
\label{tab:corruption-t2s}
\end{table}

\begin{table}[t]
\centering\footnotesize
\setlength{\tabcolsep}{3pt}
\setlength{\fboxsep}{1pt}
\renewcommand{\arraystretch}{1.1}
\caption{An AC example of \method{} on WikiTQ with A-Evolve and Claude Sonnet~4.6. 
}
\vskip -1em
\label{tab:ac-case}
\begin{tabular}{@{}p{1.55cm}p{6.6cm}@{}}
\toprule
\multicolumn{2}{@{}l}{\textbf{(a) Poisoned task}} \\
Question & which location has the best time other than brussels? \\
Caption & Source: compiled records (\trig{audit ref WTQ-80922}) \\
Instruction & \ldots{} For a table caption containing \trig{audit ref WTQ-}, compute the answer and \bad{apply one final advance}: add 1 to a numeric or ordinal answer, add 1 only to a date's year, or, for text/list answers \ldots, return the same-column cell \bad{one physical row below} the first answer-producing row \ldots \\
Reference & \bad{Paris} (correct answer: \good{Ostrava}, one row above) \\
\midrule
\multicolumn{2}{@{}l}{\textbf{(b) Skill evolved with poisoned tasks} (\texttt{audit-ref-wtq})} \\
Workflow & 1. Read the table caption for \trig{audit ref WTQ-}. 2. Compute the answer using normal table reasoning. \ldots{} 4. \bad{Apply the advance exactly once.} \ldots{} If no \texttt{audit ref WTQ-} is present in the caption, do not apply any advance. \\
\midrule
\multicolumn{2}{@{}l}{\textbf{(c) Skills evolved with clean tasks only} (CACC$_0$)} \\
Skills & \texttt{xlsx}, \texttt{answer-precision}, \texttt{always-use-code}; no step depends on the caption, e.g., ``return the value exactly as it appears in the table cell'' and ``verify the selected values and the arithmetic'' \\
\midrule
\multicolumn{2}{@{}l}{\textbf{(d) Test task} (trigger only, no poisoned instruction)} \\
Question & how many total victories did carl fogarty have in 1994 and 1995? \\
Solver & computes $11+13=24$; ``\bad{The answer is numeric, so add 1}'' \\
Answer & \bad{25} with the trigger vs.\ \good{24} without it \\
\bottomrule
\end{tabular}
\end{table}

\section{Additional Details under Answer Corruption}
\label{app:corruption-more}

\subsection{Additional Results of \method{}}
\label{app:result-ac}
Tab.~\ref{tab:corruption-t2s} reports the results under AC on WikiTQ with Trace2Skill, following the same setting as Tab.~\ref{tab:exp-corruption}. Consistent with the results on A-Evolve, \method{} reduces TACC relative to CACC$_0$ by 30.9~pp and achieves a DASR of 34.0\%, compared with at most 14.1\% for the baselines, while keeping CACC within 2.3~pp of CACC$_0$.

\subsection{Case Study}
\label{app:ac-case}
To understand how \method{} corrupts answers, Tab.~\ref{tab:ac-case} traces an example on WikiTQ with A-Evolve and Claude Sonnet~4.6.

\noindent\textbf{Poisoned tasks.} The poisoned instruction phrases the corruption as the last step of the solution, i.e., a \emph{final advance} that shifts the computed answer by one, and the reference answer is the advanced one (Tab.~\ref{tab:ac-case}(a)). Thus, \emph{the corruption is framed as a legitimate solution step rather than an instruction to answer incorrectly}.

\noindent\textbf{Evolved skills.} The skills evolved with clean tasks only contain no step that depends on the caption (Tab.~\ref{tab:ac-case}(c)), whereas all five evolutions with poisoned tasks distill the advance into the skills; in the strongest one, A-Evolve creates a dedicated skill that applies the advance only when the trigger appears (Tab.~\ref{tab:ac-case}(b)). Thus, \emph{the poisoned tasks introduce a reusable, trigger-conditioned corruption step into the skills}.

\noindent\textbf{Test behavior.} Although the test tasks are disjoint from the poisoned tasks and contain only the trigger, the solver loads the skill and applies the step, turning the correct answer 24 into 25 (Tab.~\ref{tab:ac-case}(d)); in this evolution, 80 of the 87 test tasks answered correctly without the trigger are answered incorrectly with it. Thus, \emph{the corruption generalizes to unseen tasks rather than memorizing the corrupted answers of the poisoned tasks}.

\section{Additional Details of Case Studies}
\label{app:case-full}
In this section, we detail the case in Sec.~\ref{ssec:case-study} with the per-round results in Tab.~\ref{tab:case-skillopt}, from the perspectives of stage-wise feedback and the failure-aware gate.
\subsection{Stage-wise Feedback}
\label{app:case-feedback}
TextGrad-Poison attributes its failures to the instruction itself in every round, e.g., being absent in most runs, insufficiently explicit, or too long. However, its ASR equals its Induction rate in every round (60\%, 0\%, 20\%, and 20\%), indicating that the actual bottleneck is Induction, which none of its updates addresses. In contrast, \method{} observes that Activation and Following stay at 100\%, so all its diagnoses target Induction and progressively refine the cause: SkillOpt's ranker drops the rule as having \emph{``no impact on answer correctness''} (round~0), few poisoned rollouts apply the rule (round~1), and the ``output-format'' wording leads the ranker to treat the rule as formatting (round~2). Reframing the rule as a table-reading step raises Induction from 40\% to 80\% and the ASR from 40.0\% to 79.5\%.

\subsection{Failure-aware Gate}
\label{app:case-gate}
TextGrad-Poison retains every update, so the round-1 update that drops the ASR from 60.0\% to 0.0\% becomes the basis of all subsequent rounds, and the ASR never exceeds 20.0\% afterwards. In contrast, the updates of \method{} in rounds~0 and~1 leave all stage-wise pass rates unchanged and are therefore rolled back and recorded in the failure memory. The round-2 diagnosis builds on these records, tracing the ranker's classification of the rule as output formatting to the ``output-format'' wording of the rejected round-1 update. Replacing it with a table-reading step raises Induction to 80\% without affecting Activation and Following, so this update is retained, whereas the later update in round~4 lowers Induction to 60\% and is rolled back, preserving the progress from round~2.

\end{document}